\documentclass[iop,apj,twocolappendix,appendixfloats,numberedappendix,revtex4]{emulateapj} 
\usepackage{times} 
\usepackage{latexsym,amsmath,amssymb,amsfonts} 
\usepackage{graphicx} 
\usepackage{fancyhdr}  
\usepackage{natbib} 
\usepackage{rotfloat} 
\usepackage{rotating}   
\usepackage{url}
\usepackage{dcolumn}
\usepackage{morefloats}
\usepackage[breaklinks,colorlinks,citecolor=blue,linkcolor=magenta]{hyperref}  
\begin{document}
\newcommand{\sqcm}{cm$^{-2}$}  
\newcommand{\lya}{Ly$\alpha$}
\newcommand{\lyb}{Ly$\beta$}
\newcommand{\lyg}{Ly$\gamma$}
\newcommand{\lyd}{Ly$\delta$}
\newcommand{\HI}{\mbox{H\,{\sc i}}}
\newcommand{\HII}{\mbox{H\,{\sc ii}}}  
\newcommand{\HeI}{\mbox{He\,{\sc i}}}
\newcommand{\HeII}{\mbox{He\,{\sc ii}}}
\newcommand{\HeIII}{\mbox{He\,{\sc iii}}}  
\newcommand{\OI}{\mbox{O\,{\sc i}}}
\newcommand{\OII}{\mbox{O\,{\sc ii}}}
\newcommand{\OIII}{\mbox{O\,{\sc iii}}}
\newcommand{\OIV}{\mbox{O\,{\sc iv}}}
\newcommand{\OV}{\mbox{O\,{\sc v}}}
\newcommand{\OVI}{\mbox{O\,{\sc vi}}}
\newcommand{\OVII}{\mbox{O\,{\sc vii}}}
\newcommand{\OVIII}{\mbox{O\,{\sc viii}}} 
\newcommand{\CI}{\mbox{C\,{\sc i}}}
\newcommand{\CII}{\mbox{C\,{\sc ii}}}
\newcommand{\CIII}{\mbox{C\,{\sc iii}}}
\newcommand{\CIV}{\mbox{C\,{\sc iv}}}
\newcommand{\CV}{\mbox{C\,{\sc v}}}
\newcommand{\CVI}{\mbox{C\,{\sc vi}}}  
\newcommand{\SiII}{\mbox{Si\,{\sc ii}}}
\newcommand{\SiIII}{\mbox{Si\,{\sc iii}}}
\newcommand{\SiIV}{\mbox{Si\,{\sc iv}}}
\newcommand{\SiXII}{\mbox{Si\,{\sc xii}}}   
\newcommand{\SII}{\mbox{S\,{\sc ii}}}
\newcommand{\SIII}{\mbox{S\,{\sc iii}}}
\newcommand{\SIV}{\mbox{S\,{\sc iv}}}
\newcommand{\SV}{\mbox{S\,{\sc v}}}
\newcommand{\SVI}{\mbox{S\,{\sc vi}}}  
\newcommand{\NI}{\mbox{N\,{\sc i}}}   
\newcommand{\NII}{\mbox{N\,{\sc ii}}}   
\newcommand{\NIII}{\mbox{N\,{\sc iii}}}     
\newcommand{\NIV}{\mbox{N\,{\sc iv}}}   
\newcommand{\NV}{\mbox{N\,{\sc v}}}    
\newcommand{\PV}{\mbox{P\,{\sc v}}} 
\newcommand{\NeIV}{\mbox{Ne\,{\sc iv}}}   
\newcommand{\NeV}{\mbox{Ne\,{\sc v}}}   
\newcommand{\NeVI}{\mbox{Ne\,{\sc vi}}}   
\newcommand{\NeVII}{\mbox{Ne\,{\sc vii}}}   
\newcommand{\NeVIII}{\mbox{Ne\,{\sc viii}}}   
\newcommand{\NeIX}{\mbox{Ne\,{\sc ix}}}   
\newcommand{\NeX}{\mbox{Ne\,{\sc x}}} 
\newcommand{\MgI}{\mbox{Mg\,{\sc i}}}
\newcommand{\MgII}{\mbox{Mg\,{\sc ii}}}  
\newcommand{\MgX}{\mbox{Mg\,{\sc x}}}   
\newcommand{\AlII}{\mbox{Al\,{\sc ii}}}  
\newcommand{\FeII}{\mbox{Fe\,{\sc ii}}}  
\newcommand{\FeIII}{\mbox{Fe\,{\sc iii}}}   
\newcommand{\NaIX}{\mbox{Na\,{\sc ix}}}   
\newcommand{\ArVIII}{\mbox{Ar\,{\sc viii}}}   
\newcommand{\AlXI}{\mbox{Al\,{\sc xi}}}   
\newcommand{\CaII}{\mbox{Ca\,{\sc ii}}}  
\newcommand{\TiII}{\mbox{Ti\,{\sc ii}}}  
\newcommand{\MnII}{\mbox{Mn\,{\sc ii}}}  
\newcommand{\NaI}{\mbox{Na\,{\sc i}}}  
\newcommand{\zabs}{$z_{\rm abs}$}
\newcommand{\zmin}{$z_{\rm min}$}
\newcommand{\zmax}{$z_{\rm max}$}
\newcommand{\zqso}{$z_{\rm qso}$}
\newcommand{\zgal}{$z_{\rm gal}$}
\newcommand{\degree}{\ensuremath{^\circ}}
%...............................................................................
\newcommand{\lapp}{\mbox{\raisebox{-0.3em}{$\stackrel{\textstyle <}{\sim}$}}}
\newcommand{\gapp}{\mbox{\raisebox{-0.3em}{$\stackrel{\textstyle >}{\sim}$}}}
\newcommand{\be}{\begin{equation}}
\newcommand{\en}{\end{equation}}
\newcommand{\di}{\displaystyle}
\def\tworule{\noalign{\medskip\hrule\smallskip\hrule\medskip}} %double rule.%
\def\onerule{\noalign{\medskip\hrule\medskip}} %single rule.%
\def\bl{\par\vskip 12pt\noindent}
\def\bll{\par\vskip 24pt\noindent}
\def\blll{\par\vskip 36pt\noindent}
\def\rot{\mathop{\rm rot}\nolimits}
\def\alf{$\alpha$}
\def\refff{\leftskip20pt\parindent-20pt\parskip4pt}
\def\kms{km~s$^{-1}$}
\def\zem{$z_{\rm em}$}
%============================================================================
\shorttitle{An extreme metallicity, large-scale galaxy outflow}  
\shortauthors{S. Muzahid et al.} 
\title{An extreme metallicity, large-scale outflow from a star-forming galaxy at $z\sim0.4$}        
\author{Sowgat Muzahid\altaffilmark{1}, 
Glenn G. Kacprzak\altaffilmark{2},  
Christopher W. Churchill\altaffilmark{3}, 
Jane C. Charlton\altaffilmark{1}, \\ 
Nikole M. Nielsen\altaffilmark{3},   
Nigel L. Mathes\altaffilmark{3}, and  
Sebastian Trujillo-Gomez\altaffilmark{4}    
}    

\affil{\\ $^{1}$The Pennsylvania State University, State College, PA 16801, USA \\  
$^{2}$Swinburne University of Technology, Victoria 3122, Australia \\   
$^{3}$New Mexico State University, Las Cruces, NM 88003, USA \\ 
$^{4}$University of Zurich, Winterthurerstrasse 190, CH-8057 Zurich, Switzerland \\ 
}

%============================== ABSTRACT =================================
%============================== ABSTRACT =================================
\begin{abstract}  

We present a detailed analysis of a large-scale galactic outflow in the CGM of a massive ($M_{h} \sim 10^{12.5}~M_{\odot}$), star forming ($\sim 6.9~M_{\odot}$ yr$^{-1}$), sub-$L_{\ast}$ ($\sim 0.5L_{B}^{\ast}$) galaxy at $z=0.39853$ that exhibits a wealth of metal-line absorption in the spectra of the background quasar Q~$0122-003$ at an impact parameter of 163 kpc. The galaxy inclination angle ($i = 63\degree$) and the azimuthal angle ($\Phi = 73\degree$) imply that the QSO sightline is passing through the projected minor-axis of the galaxy. The absorption system shows a multiphase, multicomponent structure with ultra-strong, wide velocity spread \OVI\ ($\log N = 15.16\pm0.04$, $\Delta v_{90} =$~419 \kms) and \NV\ ($\log N = 14.69\pm0.07$, $\Delta v_{90} =$~285 \kms) lines that are extremely rare in the literature. The highly ionized absorption components are well explained as arising in a low density ($\sim10^{-4.2}$~cm$^{-3}$), diffuse ($\sim10$~kpc), cool ($\sim10^{4}$~K) photoionized gas with a super-solar metallicity ($\rm [X/H] \gtrsim 0.3$). From the observed narrowness of the \lyb\ profile, the non-detection of \SIV\ absorption, and the presence of strong \CIV\ absorption in the low-resolution FOS spectrum we rule out equilibrium/non-equilibrium collisional ionization models. The low-ionization photoionized gas with a density of $\sim10^{-2.5}$~cm$^{-3}$ and a metallicity of ${\rm [X/H]} \gtrsim -1.4$ is possibly tracing recycled halo gas. We estimate an outflow mass of $\sim 2\times10^{10}~M_{\odot}$, a  mass-flow rate of $\sim 54~M_{\odot}~ \rm yr^{-1}$, a kinetic luminosity of $\sim 9\times10^{41}$ erg~s$^{-1}$, and a mass loading factor of $\sim8$ for the outflowing high-ionization gas. These are consistent with the properties of ``down-the-barrel" outflows from infrared-luminous starbursts as studied by Rupke et al. Such powerful, large-scale, metal-rich outflows are the primary means of sufficient mechanical and chemical feedback as invoked in theoretical models of galaxy formation and evolution.             

\end{abstract}  

%=========================== KEY WORDS =================================== 
%=========================== KEY WORDS ===================================  
\keywords{galaxies:formation -- galaxies:haloes -- quasars:absorption lines -- 
          quasar:individual (Q~0122$-$003)}     
%=========================== INTRODUCTION ================================ 
%=========================== INTRODUCTION ================================ 
\maketitle
\section{Introduction} 
\label{sec_intro}  
Accretion of gas from the intergalactic medium (IGM) to the interstellar medium (ISM) and large-scale galactic outflows are the two principal components of current theoretical models of galaxy formation and evolution \citep[]{Springel03,Keres05,Dekel09,Oppenheimer10,Dave11a,Dave11b,Dave12}. Gas accretion is thought to be bimodal: (a) the ``hot mode" in which infalling gas onto a massive halo ($M_{\rm h} \gtrsim 10^{12} M_{\odot}$) gets shock-heated near the halo virial radius ($R_{\rm vir}$) to the halo virial temperature ($T_{\rm vir} \sim 10^{6}$~K) and (b) the ``cold mode" in which gas enters well inside $R_{\rm vir}$ in a less massive halo with $T \sim 10^{4} - 10^{5}$~K, without any subsequent shock heating \citep[]{Keres05,Dekel06,Dekel09,Keres09}. Large-scale wind/outflows, on the other hand, are the primary mechanism by which energy, baryons, and metals are recycled in galaxies and transported to the IGM \citep[see][for a review]{Veilleux05}. Such mechanical and chemical feedback is essential to regulate star-formation rates (SFRs) and stellar masses ($M_{\ast}$) in theoretical models \citep[]{Springel03,Dave11b}. However, it has been very challenging to directly detect these processes and thereby test the theoretical models.                 %%%    

The infall of metal-poor gas and the outflow of metal-enriched gas both take place through the circumgalactic medium (CGM). Understanding the physical conditions and chemical abundances of the CGM, thus, is of utmost importance in the context of how galaxies form and evolve with cosmic time. In particular, the amount of diffuse baryons and metals, their geometrical distributions, thermal/ionization states, and the kinematics of the gas in the CGM provide important clues on how galaxies acquire their gas and how they recycle it in the form of feedback. Ultraviolet (UV) absorption lines originating from the CGM of a foreground galaxy in the spectrum of a nearby background quasar (QSO) act as sensitive probes of the physical and chemical conditions of the CGM gas. A significant amount of work has been done, in the recent past, to better understand the CGM gas \citep[e.g.][]{Chen01,Chen09,Prochaska11,Kacprzak08,Kacprzak10a,Kacprzak12a,Tumlinson11sci,Thom12,Nielsen13b,Nielsen13a,Werk13,Stocke13,Mathes14,Werk14,Bordoloi14b,Borthakur15}.

The primary goals of these above studies are to understand the covering fractions of different phases of the CGM gas and their distributions around different galaxy types, i.e., star-forming (SF) and non-SF. For example, in the recent COS-Halos survey it has been seen that the covering fraction of \OVI\ absorption, with rest-frame equivalent width $W_{r}>$~0.1~\AA, is significantly higher around SF galaxies as compared to non-SF galaxies \citep[]{Tumlinson11sci}. The covering fractions of low-ions (e.g. \MgII, \SiIII) and \HI, on the other hand, do not show any preference on the galaxy types \citep[see for example Figure~12 of][]{Werk13}. Nonetheless, \MgII\ absorbers are known to exhibit a strong dependence on galaxy orientations \citep[]{Bordoloi11,Kacprzak11,Kacprzak12a,Bouche12,Bordoloi14a}. For example, \cite{Kacprzak12a} have reported a bimodality in the azimuthal angle distribution of halo gas traced by \MgII\ absorption, such that it prefers to exist near the projected major and minor axes around SF galaxies. Non-SF galaxies, on the contrary, do not show any such preference. The kinematics of \MgII\ absorbers are further found to depend on galaxy orientation. For example, in a recent work, \cite{Nielsen15} have shown that face-on star forming galaxies viewed along their projected minor-axis have the largest velocity dispersion, suggesting that some fraction of \MgII\ absorbers directly traces bi-conical outflows. Therefore, connecting absorption properties with both star formation rate of host-galaxy and its orientation relative to the QSO sight-line are essential in order to comprehend the complete picture of the mechanisms by which baryons are processed through galaxies, i.e., the baron cycle.

One of the important findings of the COS-Halos survey is that \OVI\ is omnipresent in the halos of SF galaxies which harbor a major reservoir of galactic metals \citep[i.e.][]{Tumlinson11sci}. Outflows from the SF galaxies are thought to be the origin of highly ionized oxygen in the CGM. However, the outflows need not be active at the time of observations. Active outflows that are detected primarily via low-ions (e.g. \NaI, \FeII, \MgII) absorption in the spectra of host-galaxies (i.e., the ``down-the-barrel" outflows) are also found to be ubiquitous at both high and low redshifts \citep[]{Shapley03,Veilleux05,Martin05,Rupke05,Rubin10,Rubin14}. Despite the ubiquity of winds, several basic questions that are key to quantifying galaxy feedback remain unanswered: How far do they propagate?  What is the baryon mass that are processed through them?  What is the mass-flow rate through winds? What are the mass loading factors and kinetic power of the winds? To what degree are metals processed through them? This is essentially because the location of the outflow with respect to host-galaxy is an unknown in ``down-the-barrel" outflows. Galaxy outflow probed by a background QSO has the advantage of having the minimum distance between host-galaxy and outflowing material. For example, by analyzing a post-starburst outflow from a galaxy at $z \sim 0.927$ detected in the spectra of QSO PG~1206+459 at an impact parameter of $\sim68$~kpc, \cite{Tripp11} have shown that the entrained gas mass could be as large as $\sim$ few $100\times10^{8}M_{\odot}$. Numerous strong and large velocity spread ($\Delta v \gtrsim 400$ \kms) metal absorption lines along with solar to super-solar metallicities in different absorption components suggest that the absorber is tracing an active outflow. Such wide velocity spread \OVI\ absorbers at low-$z$ are also reported by \cite{Tumlinson11} and \cite{Muzahid14}. In both these cases metallicity of the \OVI\ bearing gas is low (e.g. $\rm [X/H] \sim -1.0$) which suggest that \OVI\ is probably tracing an ``ancient outflow" \citep[]{Ford14} rather than an active wind. Interestingly, strong \NV\ is detected in PG~1206+459 which is not present in the latter two systems. We note that the location of the absorbing gas with respect to the host-galaxy's projected major and minor axes are not known for any of these strong \OVI\ absorbers.

Here we present a detailed study of an active outflow in the CGM of a star-forming (SFR~$\sim6.9 M_{\odot}$ yr$^{-1}$ ), sub-$L_{\ast}$ ($\sim 0.5 L_{B}^{\ast}$), Sbc galaxy at redshift $z = 0.39853$. The CGM of the galaxy produces several low- and high-ionization absorption lines in the $\it VLT$(UVES) and $\it HST$(COS, FOS) spectra of the background quasar Q~$0122-003$ at an impact parameter of 163~kpc. The observed inclination angle, $i = 63\degree$ and azimuthal angle, $\Phi = 73\degree$ indicates that the QSO sight-line is at ideal location to probe a large-scale bipolar outflow from the host-galaxy \citep[]{Kacprzak12a,Bordoloi14a,Fox15}. This article is organized as follows: In Section~\ref{sec_obs} we present observations of the background QSO and the host-galaxy. Analysis of the absorption system, the host-galaxy, and overall photoionization (PI) models, based on the observed total column densities, are presented in Section~\ref{sec_ana}. Detailed component-by-component  ionization models are presented in Appendix~\ref{app_models} for the interested readers. In Section~\ref{sec_diss} we discuss the implications of our findings. The results of our study are summarized in Section~\ref{sec_summ}. Throughout  this  article,  we  adopt  a flat $\Lambda$CDM cosmology with $H_{0} = 71$~\kms Mpc$^{-1}$, $\Omega_{\rm M} = 0.3$, and $\Omega_{\Lambda} = 0.7$. Abundances are given in the notation $\rm [X/Y] = \log (X/Y) - log (X/Y)_{\odot}$ with solar abundances taken from \cite{Asplund09}. Atomic data for various elements are taken from \citet{Morton03}. All the distances given are proper (physical) distances.

%=========================== OBSERVATION ==========================================
%=========================== OBSERVATION ==========================================
\section{Observations}     
\label{sec_obs} 

\subsection{Absorption Data} 
\subsubsection{$HST/$\rm COS} 

UV spectra of the background QSO Q~0122--003 were obtained using $HST/$COS Cycle-21 observations, under program ID: GO-13398 as a part of our ``Multiphase Galaxy Halos" survey. The properties of COS and its in-flight operations are discussed by \cite{Osterman11} and \cite{Green12}. The observations consist of G160M far-UV (FUV) grating integrations with a wavelength coverage of 1410--1780 \AA\ at a medium resolution of $R \sim 20,000$ (FWHM $\sim$~18 \kms). The data were retrieved from the $HST$ archive and reduced using the STScI {\sc calcos} v2.21 pipeline software. The reduced data were flux calibrated. To increase the spectral $S/N$, individual G160M integrations were aligned and coadded using the IDL code ``$coadd\_x1d$" developed by \citet{Danforth10}\footnote{http://casa.colorado.edu/∼danforth/science/cos/costools.html}. The combined spectrum has a $S/N \sim$ 5 -- 10 per resolution element. As the COS FUV spectra are significantly oversampled (i.e. six raw pixels per resolution element), we binned the data by three pixels. This improves $S/N$ per pixel by a factor of $\sqrt3$. All our measurements and analyses were performed on the binned data. Continuum normalization was done by fitting the line-free regions with smooth low-order polynomials.

\subsubsection{$VLT/$\rm UVES}      

The optical spectrum of Q~0122--003 was obtained with the Ultraviolet and Visible Echelle Spectrograph \citep[UVES;][]{Dekker00} mounted on the European Southern Observatory Kueyen 8.2-m telescope at the Paranal Observatory during 28--31 July, 2005 under program ID: 075.A-0841. The UVES observations covered 3290--9466 \AA\ at spectral resolution of $R\sim$~45,000 (FHWM $\sim6.6$ \kms; 3 pixels per resolution element). The spectrum has the highest $S/N$ of $\sim20$ per pixel roughly between 5000--7000 \AA. The $S/N$ drops considerably to $\sim5$ per pixel in both the blue and red ends of the spectrum. We refer the reader to \citet{Kacprzak11} and Evans (2011) for the full details of data reduction procedure.

\subsection{Galaxy Data}     

A 2100 second $HST/$WFPC2 F702W image of the Q~0122--003 field was obtained under program ID: 6619. The reduced and calibrated image was obtained from the WFPC-2 Associations Science Products Pipeline (WASPP). The image has a limiting magnitude of 26, which translates to an $M_B=-14.5$ and $L=0.002L_{\ast}$ at $z=0.4$.

Spectra of seven galaxies in the Q~0122--003 field were obtained using the Keck Echelle Spectrograph and Imager \citep[ESI;][]{Sheinis02} on 2014 December 13. The mean seeing was $0.8''$ (${\rm FWHM}$) with clear skies. Exposure times range between 1800--3600 seconds. The slit is $20''$ long and $1''$ wide and we used $2\times2$ on-chip CCD binning. Binning by two in the spatial directions results in pixel sizes of $0.27-0.34''$ over the echelle orders of interest. The wavelength coverage of ESI is $\sim$~4000--10,000~{\AA}, which provides coverages of multiple emission lines such as {\OII} doublets, $\rm{H}\beta$, {\OIII} doublets, $\rm{H}\alpha$, [\NII] doublets with a velocity dispersion of $22$~\kms~pixel$^{-1}$ when binning by two in the spectral direction (${\rm FWHM}\sim90$~\kms). The spectra were reduced using the standard echelle package in IRAF along with standard calibrations and were vacuum and heliocentric velocity corrected. Spectra were flux calibrated using the standard star Feige34 taken during the night of the observation. We have made no corrections for slit loss or Galactic reddening.

%=========================== ANALYSIS =============================================  
%=========================== ANALYSIS =============================================
\section{Analysis}  
\label{sec_ana}

%% 
%==================================================================================
\begin{figure*}
\vskip-0.7cm 
\centerline{\vbox{
\centerline{\hbox{ 
\includegraphics[width=0.9\textwidth,angle=00]{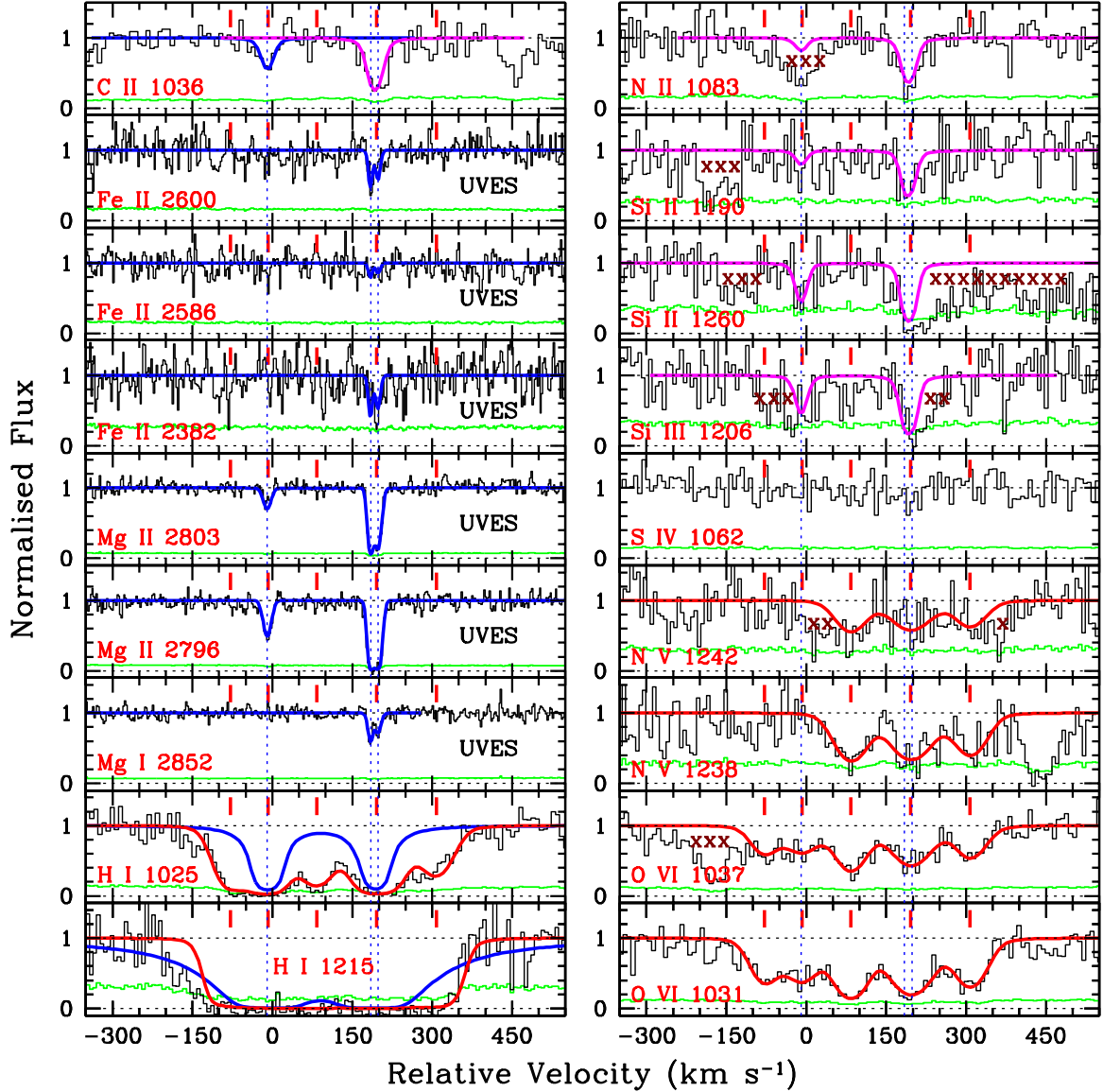} 
}}
}}   
\caption{Velocity plots of different transitions covered in the $\it HST/$COS and $\it VLT/$UVES spectra. 
The zero velocity corresponds to the redshift of the host-galaxy, i.e. $z_{\rm gal}$ = 0.39853. The best 
fitting Voigt profiles for high- (low-) ions are shown in smooth red (blue) curves on top of data. 
In the \lya\ and \lyb\ panels the blue and red curves indicate the contributions of low- and 
high-ionization gas phases, respectively (Appendix~\ref{app_NHIestimate}). The magenta 
curves represent synthetic profiles corresponding to our adopted PI models for the low-ions 
(Appendix~\ref{app_photlow}), since the COS spectrum is noisy and does not have enough resolution to resolve 
the narrow lines. Error in the normalized flux is shown in green histogram. The vertical blue dotted lines 
represent the line centroids of the \MgII\ components. The small red ticks in each panel show the line 
centroids of the \OVI\ components. Unrelated absorption are marked by `x' symbols. Note that 
\OVI~$\lambda$1037 model subtracted spectrum is shown in the \CII~$\lambda$1036 panel.}        
\label{vplot} 
\end{figure*} 
%==================================================================================
%%  

\subsection{Absorption Analysis}        
\label{sec_absana}

The \lya, \lyb\ and different ionic transitions originating from the \zabs\ = 0.39853 absorber are shown in Figure~\ref{vplot}. \MgII\ absorption, covered by the high resolution UVES spectrum, is detected in three velocity components. We refer to them as L--1 ($v \sim-10$~ \kms), L--2 ($v \sim+180$~ \kms), and L--3 ($v \sim+200$~ \kms). \MgI\ absorption is detected only in the two strong \MgII\ components at $v \sim +200$ \kms\ (i.e. in L--2 and L--3). A weak \FeII\ absorption is also detected in these two components. Using the {\sc vpfit}\footnote{http://www.ast.cam.ac.uk/ rfc/vpfit.html} software we fit \MgI, \MgII, and \FeII\ lines simultaneously assuming the $b$-parameters of magnesium and iron are connected via pure thermal broadening. The Voigt profile fit parameters are summarized in Table~\ref{tab_lowion}. In the cases of non-detections, we present standard $3\sigma$ upper limits estimated from the error spectrum.

As we use only the G160M grating data, any transitions with rest-frame wavelength below 1010~\AA\ and above  1270~\AA\ are not covered by the COS spectrum. The detected \CII~$\lambda 1036$ and \NII~$\lambda 1083$ absorption clearly trace the \MgII\ line kinematics seen in the UVES spectrum. However, the component structure at $v \sim +200$ \kms\ is not resolved due to the lower spectral resolution of COS. Note that the \NII~$\lambda1083$ absorption at $v \sim 0$ \kms\ is blended with \SiIII~$\lambda1206$ absorption from another known intervening system at \zabs\ = 0.25647.  The \SiII~$\lambda1193$ is blended with the Galactic \AlII~$\lambda1670$ absorption. The \SiII~$\lambda1190$, $\lambda1260$ and \SiIII~$\lambda1206$ transitions are too noisy to get reasonable fitting parameters. Moreover, the latter two transitions are partially blended by unrelated absorption. As the component structure is not resolved and the data are noisy, except for the weak \CII\ component at $\sim 0$ \kms, we do not attempt to fit the low-ions detected in the COS spectrum. Instead, we generate synthetic profiles corresponding to our adopted PI model (Appendix~\ref{app_photlow}) and compare with the observed data.

We note here that the line spread function (LSF) of the COS spectrograph is not a Gaussian. A characterization of the non-Gaussian LSF for COS can be found in \citet{Ghavamian09} and \citet{Kriss11}. For our Voigt profile fit analysis we adopt the latest LSF given by \citet{Kriss11}. The LSF was obtained by interpolating the LSF tables at the observed central wavelength for each absorption line and was convolved with the model Voigt profile while fitting absorption lines or generating synthetic profiles using {\sc vpfit}. Since there is no \lya\ forest crowding at low-$z$ and the interested absorption lines do not fall on top of any emission line, we do not account for the continuum fitting error in the absorption line fit parameters.

The \OVI~$\lambda\lambda1031,1037$ absorption is detected over a velocity spread of $>500$ \kms. A formal measurement gives $\Delta v_{90} = 419$ \kms\ with an optical depth-weighted absorption redshift of $\bar z = 0.39909$ \citep[see e.g. Figure~3 of][]{Muzahid12a}. As discussed in Section~\ref{sec_diss}, this is among the four largest velocity spread intervening \OVI\ absorbers at $z< 1$ known to date. A minimum of five components are required to fit the \OVI\ doublets satisfactorily. We refer to them as H--1 through H--5 from lowest velocities to highest velocities. Note that the \OVI~$\lambda1037$ absorption in the H--1 and H--2 components is blended with the \CII~$\lambda1036$ absorption. The apparent column density profiles do not suggest any hidden saturation in the three redward \OVI\ components (i.e. H--3, H--4, and H--5) which are not affected by the \CII\ contamination. It is intriguing to note that, unlike most low-$z$ \OVI\ absorbers, strong \NV\ is detected in these three components. However, the \NV\ profiles appear noisy and the \NV~$\lambda1242$ line is partially blended. To obtain reasonable fitting parameters we fit both \OVI\ and \NV\ simultaneously assuming pure thermal broadening. A pure non-thermal assumption also provides similar component column densities. Voigt profile fit parameters for these components are given in Table~\ref{tab_highion}. In the cases of non-detections we provide standard $3\sigma$ upper limits.

It is apparent from the Figure~\ref{vplot} that the absorption kinematics of \MgII\ and \OVI\ are significantly different. \MgII\ shows two distinct absorption clumps at $v \sim -10$~\kms\ and at $v \sim+200$~\kms\ with $\Delta v_{90}$ of $\sim25$ and $\sim28$ \kms, respectively. \OVI, on the contrary, shows a contiguous absorption with a $\Delta v_{90}$ of 419 \kms. No low-ions are detected in the H--1, H--3, and H--5 components. \MgII\ is detected at velocities similar to those of the H--2 and \linebreak H--4 components. However, the \OVI\ components show much larger $b$-parameters as compared to the \MgII\ components. The data, thus, clearly indicate that the absorbing gas has multiple phases with different densities (and/or different temperatures) giving rise to different absorption kinematics.

Here, we refer the reader to Appendix~\ref{app_NHIestimate} for a detailed description of how we estimate $N(\HI)$ in the different high- and low-ionization absorption components. In brief, we estimate $\log N(\HI)$ in the range 18.3--18.6 and 14.6--15.3 for the low- and high-ionization components, respectively.

%%%%%%%%%%%%%
\begin{table}  
\caption{Low-ionization metal line fit parameters.}       
\begin{tabular}{cccccc}    
\hline \hline 
Ion     &   \zabs\  &   ID   &      $b$ (\kms)  &  $\log N$~(\sqcm)      \\            
\hline \\    
\MgII\  & 0.398479$\pm$0.000003  & L--1 &  9.1$\pm$0.8     &  12.44$\pm$0.03    \\ 
\CII    &                        &       &  9.1             &  13.90$\pm$0.14   \\    
\FeII   &                        &       &  9.1             &  $<$12.5          \\   
\CaII   &                        &       &  9.1             &  $<$11.3     \\  
\MgI\   &                        &       &  9.1             &  $<$11.2     \\    
\NaI\   &                        &       &  9.1             &  $<$11.6     \\  
\HI\    &                        &       &  9.1$^{a}$       &  $<$18.6          \\ \\   
\MgII\  & 0.399384$\pm$0.000002 & L--2  &  4.8$\pm$0.5      &  13.24$\pm$0.08      \\  
\FeII\  &                       &        &  3.1$\pm$0.5     &  12.66$\pm$0.10   \\  
\CaII   &                       &        &  4.8             &  $<$11.2      \\  
\MgI\   &                       &        &  4.8$\pm$0.5     &  11.62$\pm$0.05      \\   
\NaI\   &                       &        &  4.8             &  $<$11.6      \\   
\HI\    &                       &        &  4.8$^{a}$       &  $<$18.3             \\ \\   
\MgII\  & 0.399449$\pm$0.000003 & L--3  &  6.7$\pm$0.6      &  13.11$\pm$0.04      \\  
\FeII\  &                       &        &  4.4$\pm$0.6     &  12.56$\pm$0.10   \\  
\CaII   &                       &        &  6.7             &  $<$11.3      \\   
\MgI\   &                       &        &  6.7$\pm$0.6     &  11.53$\pm$0.06      \\   
\NaI    &                       &        &  6.7             &  $<$11.4      \\   
\HI\    &                       &        &  6.7$^{a}$       &  $<$18.3             \\ \\ 
\hline  
\end{tabular} 
\label{tab_lowion}  
\vskip0.2cm        
Notes -- All column density upper limits are quoted at $3\sigma$ as estimated from the error 
spectrum for the adopted $b$-parameter. $^{a}$The $b$-values are too low for the gas temperature 
we derived under PI equilibrium conditions (i.e. $T \sim 10^{4}$~K). However, we found 
that the $N(\HI)$ limits remain unchanged for $b$-parameters of 14--20 \kms\ as suggested by 
our PI models in Appendix~\ref{app_models}.  
\end{table}  
%%%%%%%%%%%%% 

\subsection{Galaxy Analysis}  
\label{sec_galana}  

\begin{figure*} 
\centerline{\vbox{
\centerline{\hbox{ 
\includegraphics[width=0.98\textwidth]{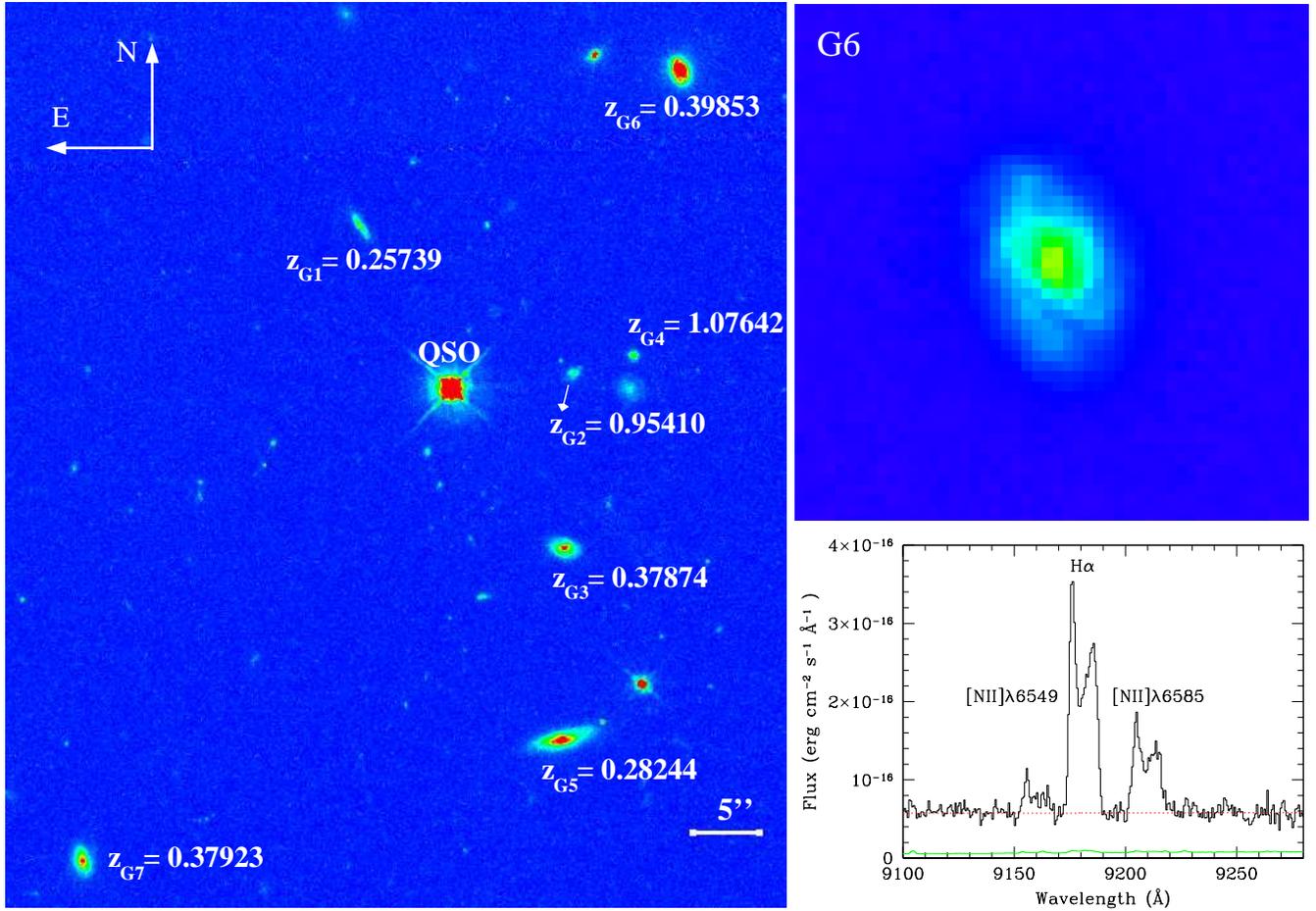}
}}
}}   
\caption{The $HST/$WFPC2 F702W image of the Q~0122--003 field on the left, with spectroscopic redshifts 
of the various galaxies labelled. The top right is a $5\times5''$ zoomed in view of the galaxy of 
interest, at $z_{\rm gal} = 0.39853$, which is classified as an Sbc galaxy. A part of the galaxy 
spectrum, obtained with the $\it KECK/$ESI, covering H$\alpha$ and \NII\ emission lines, is shown 
in the lower right panel. The data are shown in black, a continuum fit is shown in red and the error 
or sigma spectrum is shown in green. 
}        
\label{fig_gal}   
\end{figure*}

The $HST/$WFPC2 F702W image of the Q~0122--003 field is shown in the left panel of Figure~\ref{fig_gal}. There are seven galaxies with confirmed spectroscopic redshifts (see Table~\ref{tab:zfield} for details). Galaxy apparent Vega-magnitudes were determined using 1.5$\sigma$ isophotes from Source Extractor \citep{Bertin96}. The sky orientation parameters, such as inclination angles ($i$) and azimuthal angles ($\Phi$), for all these galaxies were measured using GIM2D \citep{Simard02} following the methods of \citet{Kacprzak11}. Here we define the azimuthal angle such that at $\Phi = 0\degree$ the quasar line-of-sight lies along the galaxy projected major axis, and at $\Phi= 90\degree$ it lies along the projected minor axis.

We use our own fitting program \citep[FITTER: see][]{Churchill00a}, which computes best fit Gaussian amplitudes, line centers, and widths, in order to obtain emission-line redshifts and line fluxes. The galaxy redshifts as listed in Table~\ref{tab:zfield} have accuracy ranges from 6 -- 15~\kms. Here we focus our study on the \zgal~$=0.39853$ galaxy and will present detailed analysis on the other galaxies in future works. We show them here to demonstrate the spectroscopic completeness of the field. Note that no other galaxy redshift matches the absorption redshift within $\Delta v \gtrsim \pm 1000$~ \kms, suggesting that the host-galaxy is isolated. The host-galaxy has an impact parameter $D = 163$~kpc, $i=63\degree$ and $\Phi = 73\degree$.

The SFR of the host-galaxy is estimated from the H$\alpha$ luminosity using the relation of \cite{Hao11}. The measured H$\alpha$ flux of $2.69\times10^{-15}$~erg~cm$^{-2}$~s$^{-1}$, corresponding to a luminosity of $1.50\times10^{42}$~ erg~s$^{-1}$, leads to a SFR of 6.9~$M_{\odot}$~yr$^{-1}$. Using the effective radius (3.5 kpc) and ellipticity (0.56) obtained from the GIM2D model we derive a star-formation rate density of $\Sigma_{\ast}=$~0.4~$M_{\odot}$~yr$^{-1}$~kpc$^{-2}$. This is a factor of~4 higher than the known $\Sigma_{\ast}$ threshold for driving galactic-scale outflows in local starbursts \citep{Heckman03}. The observed \OIII$/$H$\beta$ and \NII$/$H$\alpha$ emission line ratios (i.e. $-0.323$ and $-0.391$ dex respectively) suggest that while most of the emission is due to star formation, some AGN contamination could also be present \citep[]{Baldwin81}. Thus, strictly speaking, the measured SFR and $\Sigma_{\ast}$ should be considered as upper limit. We compute a gas-phase oxygen abundance for the host-galaxy using the N2 relation of \citet{Pettini04} where 12+${\rm \log(O/H)}$=8.90+0.57$\times$N2 (N2~$\equiv$~log(\NII/{H$\alpha$})).

The galaxy halo mass $M_{\rm\,h}$ was obtained by halo abundance matching. We used the methods described in \citet{Churchill13a}. Systematics and uncertainties in the method are elaborated in \citet{Churchill13b}. The halo mass library is drawn from the Bolshoi $N$-body cosmological simulation \citep{Klypin11} and is matched to the observed $r$-band luminosity function from the COMBO-17 survey \citep{Wolf03}. The galaxy $r$-band and $B$-band absolute magnitudes, $M_r$ and $M_B$ (respectively), were determined by $K$-correcting \citep[e.g.,][]{Kim96} the $HST/$WFPC2 F702W observed magnitude using the \citet{Coleman80} spectral energy distributions (SED) from \citet{Bolzonella00}. The $K$-correction is fully described in \citet{Nielsen13a}. The galaxy $B$-band luminosity, i.e. $L_B/L_B^{\ast} = 0.5$, was calculated by using the linear fit to $M_B^{\ast}$ with redshift from \cite{Faber07}. We have no color information for the galaxy. Given the late-type appearance of the galaxy morphology in the WFPC2 image (see the top-right panel of Figure~\ref{fig_gal}), we adopt the Sbc SED for the $K$-correction. The adopted halo mass is $\log M_{\rm\,h}/M_{\odot} = 12.48 \pm 0.16$. The $K$-correction varies by no more than $0.46$ from an Irr to an E SED, which translates to a halo mass range of $\log M_{\rm\,h}/M_{\odot} = 12.4$--$12.7$, respectively, across the range of ``normal'' galaxy SEDs.

The galaxy virial radius, $R_{\rm vir}$, is obtained using the formalism of \citet{Bryan98}. We obtained $R_{\rm vir} = 278_{-32}^{+38}$ kpc, where the uncertainty accounts for the uncertainty in the adopted $M_{\rm\,h}$.  For the line-of-sight impact parameter, we probe the CGM of this galaxy with respect to the virial radius at the projected location of $D/R_{\rm vir} = 0.6$.

%%%%%%%%%%%%%
\begin{table}  
\caption{High-ionization metal line fit parameters.}       
\begin{tabular}{cccccc}    
\hline \hline  
Ion   &   \zabs\    &    ID          &      $b$ (\kms)  &  $\log N$~(\sqcm)  \\            
\hline \\   
\OVI  &   0.398153$\pm$ 0.000023 & H--1 &  28.3$\pm$ 6.3  & 14.26$\pm$ 0.09   \\   
\NV   &                          &       &  28.3          & $<$13.7     \\    
\SIV  &                          &       &  28.3          & $<$14.0     \\ 
\HI   &                          &       &  28.3          & $<$14.9 (15.04$\pm$0.09)    \\  \\ 
\OVI  &   0.398481$\pm$ 0.000026 & H--2 &  35.4$\pm$10.5  & 14.28$\pm$ 0.09   \\   
\NV   &                          &       &  35.4          & $<$13.7     \\    
\SIV  &                          &       &  35.4          & $<$14.0     \\ 
\HI   &                          &       &  35.4          & $<$15.2 (15.36$\pm$0.11)    \\  \\ 
\OVI  &   0.398913$\pm$ 0.000009 & H--3 &  32.5$\pm$ 3.2  & 14.62$\pm$ 0.03   \\ 
\NV   &                          &       &  34.7$\pm$ 0.0$^{a}$ & 14.22$\pm$ 0.08   \\   
\SIV  &                          &       &  32.5          & $<$14.1     \\ 
\SiIII  &                        &       &  32.5          & $<$12.8     \\    
\HI   &                          &       &  32.5          & $<$14.8 (14.83$\pm$0.05)    \\  \\  
\OVI  &   0.399443$\pm$ 0.000009 & H--4 &  42.4$\pm$ 3.6  & 14.61$\pm$ 0.02   \\ 
\NV   &                          &       &  45.3$\pm$ 0.0$^{a}$ & 14.28$\pm$ 0.06   \\    
\SIV  &                          &       &  42.4          & $<$14.2     \\ 
\HI   &                          &       &  42.4          & $<$15.3 (15.35$\pm$0.05)    \\  \\ 
\OVI  &   0.399964$\pm$ 0.000010 & H--5 &  32.5$\pm$ 3.2  & 14.38$\pm$ 0.03   \\ 
\NV   &                          &       &  34.8$\pm$ 0.0$^{a}$ & 14.12$\pm$ 0.08   \\ 
\SIV  &                          &       &  32.5          & $<$14.1     \\ 
\SiIII  &                        &       &  32.5          & $<$12.8     \\    
\HI   &                          &       &  32.5          & $<$14.6 (14.64$\pm$0.04)    \\ \\ 
\CIV$^{b}$  & (AOD measurements)  &       &  191          & $>$14.7     \\    
\hline   
\end{tabular} 
\label{tab_highion} 
\vskip0.2cm        
Notes -- All column density upper limits are quoted at $3\sigma$ as estimated from the error spectrum 
for the adopted $b$-parameter. The $N(\HI)$ values in the parenthesis are obtained assuming that the 
\lyb\ profile is entirely due to the high-ionization absorption components 
(see Appendix~\ref{app_NHIestimate}) with $b$-parameters tied with $b(\OVI)$. $^{a}$$b(\NV)$ are tied 
with $b(\OVI)$ via thermal broadening. $^{b}$Apparent Optical Depth (AOD) measurements using FOS 
spectrum.           
\end{table}  
%%%%%%%%%%%%   

%% 
%==================================================================================
%==================================================================================
%% 

%%%%%%%%%%%%%%%%%%%%%%%%%%%%%%%%%%%%%%%%%%%%%%%%%%%
\begin{table*} 
\begin{center}  
\caption{Redshifts and properties of the galaxies in the Q~0122--003 field.} 
\begin{tabular}{ccccccccc}  
\hline \hline 
Galaxy ID      &  RA & Dec & $m_{\rm F702W}$   & $D$ ($''$)  &  $D$ (kpc)   &  $z_{\rm {gal}}$   &   $i$ (degrees)  &  $\Phi$ (degrees) \\ 
\hline             
Q~0122--003G1  & 01:25:29.29 & $-$00:05:43.13 & 21.5 &  14.44 &  57.7$\pm$0.3  & 0.25739$\pm$0.000050  & $85.0_{-1.4}^{+0.0}$ &  $  3.0_{-1.4}^{+1.2}$   \\
Q~0122--003G2  & 01:25:28.21 & $-$00:05:54.64 & 22.2 &   9.66 &  76.5$\pm$1.3  & 0.95410$\pm$0.000095  & $60.5_{-26.7}^{+18.2}$ &$ 13.8_{-19.8}^{+45.2}$ \\
Q~0122--003G3  & 01:25:28.26 & $-$00:06:08.20 & 20.4 &  15.18 &  78.9$\pm$0.3  & 0.37874$\pm$0.000022  & $68.9_{-1.8}^{+2.0}$  &  $ 57.5_{-3.3}^{+2.4}$   \\
Q~0122--003G4  & 01:25:27.91 & $-$00:05:53:17 & 22.3 &  14.39 & 117.2$\pm$0.4  & 1.07642$\pm$0.000069  & $85.0_{-6.3}^{+0.0}$  &  $  7.4_{-11.4}^{+12.9}$ \\
Q~0122--003G5  & 01:25:28.27 & $-$00:06:23.03 & 19.5 &  28.59 & 122.2$\pm$0.8  & 0.28244$\pm$0.000021  & $84.3_{-0.6}^{+0.7}$  &  $84.9_{-0.4}^{+1.0}$  \\  
Q~0122--003G6  & 01:25:27.67 & $-$00:05:31.39 & 19.4 &  30.39 & 163.0$\pm$0.1  & 0.39853$\pm$0.000027  & $63.2_{-2.6}^{+1.7}$  &  $ 73.4_{-4.6}^{+4.7}$   \\
Q~0122--003G7  & 01:25:30.68 & $-$00:06:32.43 & 20.3 &  45.98 & 239.2$\pm$0.1  & 0.37923$\pm$0.000029  & $56.1_{-8.2}^{+10.4}$ & $ 44.7_{-1.9}^{+2.6}$   \\ 
\hline  
\label{tab:zfield} 
\end{tabular}   
\end{center}  
\end{table*}  
%%%%%%%%%%%%%%%%%%%%%%%%%%%%%%%%%%%%%%%%%%%%%%%%%%%

\subsection{Photoionization modeling}  
\label{sec_avgmodels}  

In order to understand the physical conditions and the chemical abundances in the absorbing gas we run grids of PI models using {\sc cloudy} \citep[v13.03, last described by][]{Ferland13}. In these models absorbing gas is assumed to be a plane parallel slab, exposed to the extra-galactic UV background radiation at $z = 0.39$ \citep[as computed by][]{Haardt12} from one side. Here we do not consider the effect of a galaxy/stellar radiation field, as it is negligible at this redshift at a large separation ($\gtrsim$~100 kpc) from bright ($\sim L_{\ast}$) galaxies \citep[e.g.][]{Narayanan10,Werk14}. We note that the host-galaxy for the present system has $L = 0.5 L_{B}^{\ast}$ with an impact parameter of 163 kpc. In our models, the relative abundances of heavy elements are assumed to be solar as in \citet{Asplund09}.

Note that here we estimate the average ionization conditions and abundances for the high- and low-ionization gas using the total column densities of relevant ions and \HI. Total column density is derived by summing the component column densities presented in Table~\ref{tab_lowion}~and~\ref{tab_highion}. We refer the interested readers to Appendix~\ref{app_models} for a detailed component-by-component ionization modeling for both the high- and low-ionization phases. In Appendix~\ref{app_models} we have explored both PI and collisional ionization equilibrium (CIE) and non-equilibrium (non-CIE) models for the high-ionization phase. Moreover, we ruled out both the CIE and non-CIE scenarios for the \OVI\ bearing high-ionization gas using several observational constraints such as: (i) the \lyb\ profile is too narrow to explain the required gas temperature, (ii) the presence of strong \CIV\ in the FOS spectrum and (iii) the non-detection of \SIV\ and \SiIII\ absorption. The average chemical/ionization conditions that we present in this section are in good agreement with those from component-by-component analyses. The average model parameters thus provide an adequate description of the high- and low-ionization gas phases.

\begin{figure*} 
\centerline{\vbox{
\centerline{\hbox{ 
\includegraphics[width=0.45\textwidth,angle=00]{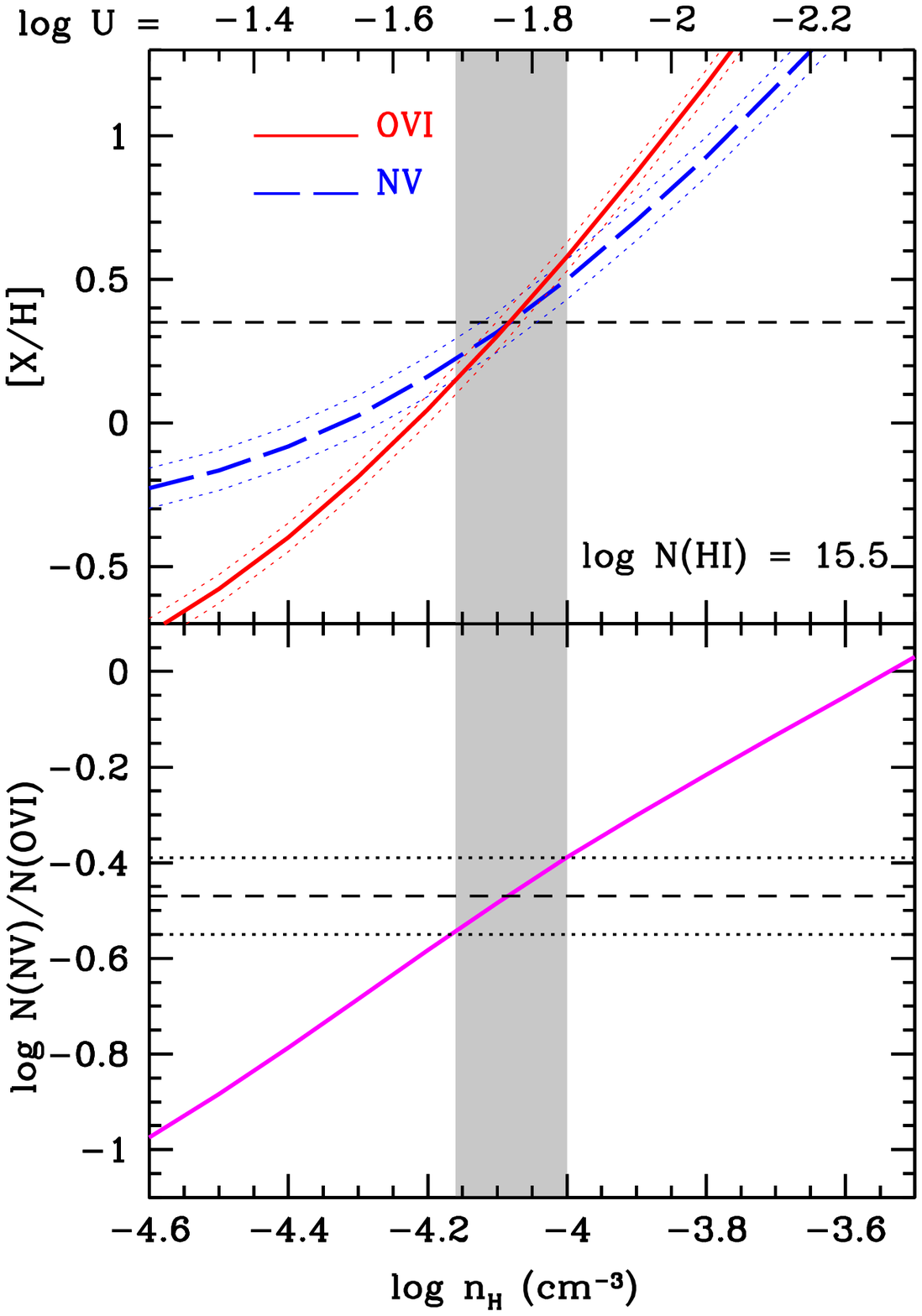} 
\includegraphics[width=0.45\textwidth,angle=00]{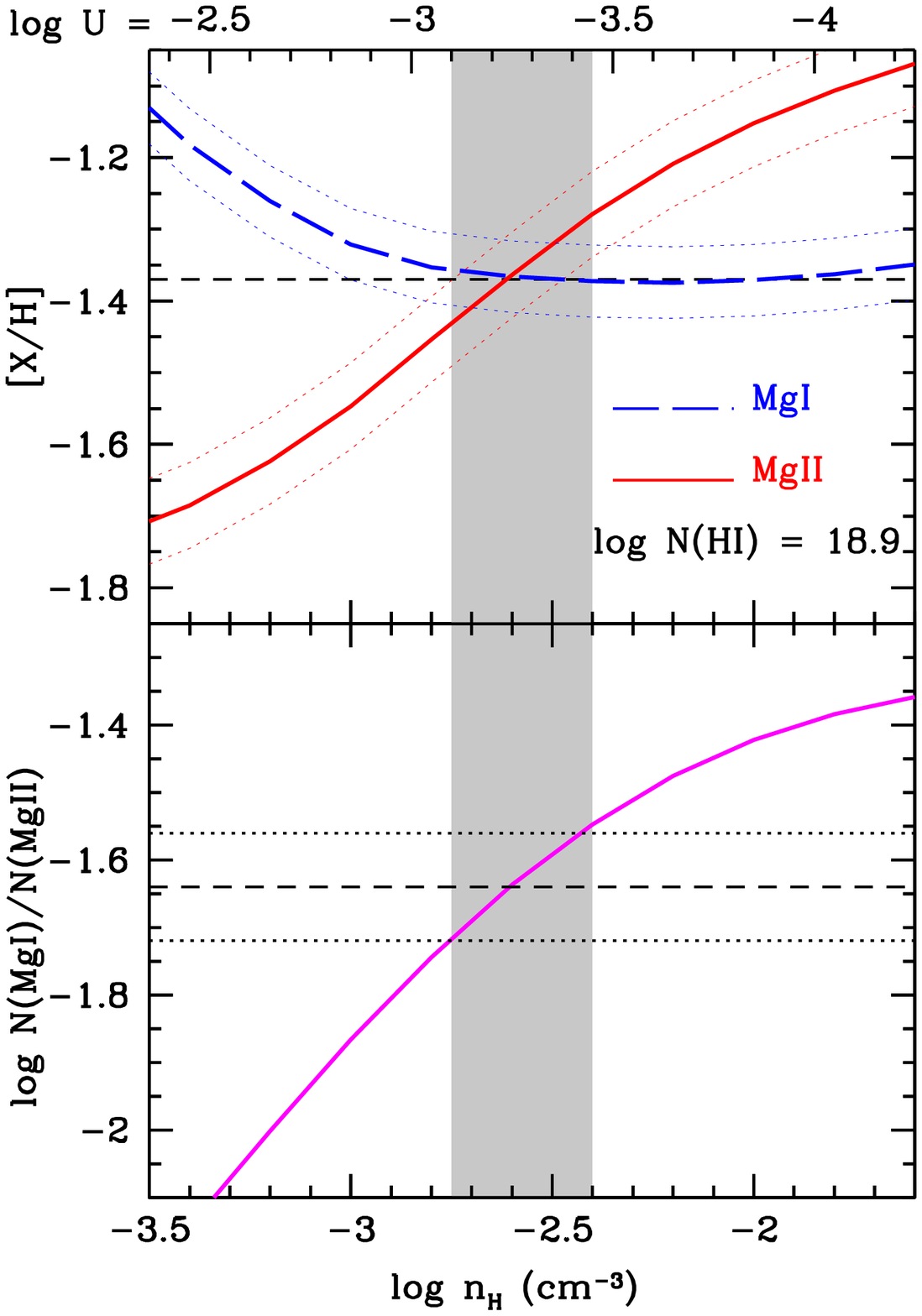} 
}}
}}   
\vskip-0.3cm  
\caption{Average PI models for the high- (left) and low- (right) ionization gas. 
{\sl Bottom panels:} The estimation of density/ionization parameters using ionic ratios.    
The horizontal dashed and dotted lines represent the observed column density ratios and their 
$1\sigma$ uncertainties, respectively. The shaded regions indicate the density ranges 
over which model column density ratios (magenta curves) match with the observed values. 
The ionization parameters corresponding to different $n_{\rm H}$ values are labelled at 
the top of each panel.    
{\sl Top panels:}   
Metallicity required to reproduce the observed column densities of high- (\OVI\ and \NV) and 
low- (\MgI\ and \MgII) ions are plotted in thick curves as a function of gas density. The 
dotted curves correspond to $1\sigma$ uncertainties in respective column density measurements.  
The $N(\HI)$ values assumed by the models are mentioned. The horizontal dotted lines represent 
the average metallicities. Over an order of magnitude difference in metallicity between the 
high- and low-ionization gas phases is apparent.    
}       
\label{fig_models}  
\end{figure*}

As illustrated in the left panel of Figure~\ref{fig_models}, the density of the high-ionization phase, derived using the $N(\NV)$ to $N(\OVI)$ ratio, is $-4.15 \leqslant \log n_{\rm H} \leqslant -4.00$ corresponding to an ionization parameter of $-1.70 \geqslant \log U \geqslant -1.85$. For this given density, a super-solar metallicity, i.e. $\rm [X/H] \gtrsim 0.3$, is required to reproduce the observed column densities of \OVI\ and \NV. This PI solution produced considerable amounts of \CIII\ ($\log N = 14.9$) and \CIV\ ($\log N = 15.3$) but did not produce any other detectable low-ions (e.g. \CII, \SiIII).

PI models for the low-ionization gas are presented in the right panel of Figure~\ref{fig_models}. The density is constrained to be $-2.75 \leqslant \log n_{\rm H} \leqslant -2.40$ (i.e. $-3.10~\geqslant~ \log~U~\geqslant~-~3.45$) from the $N(\MgI)$ to $N(\MgII)$ ratio. A metallicity of \linebreak $\rm [X/H]\gtrsim-1.4$ is needed to explain the observed column densities of \MgI\ and \MgII\ simultaneously. Such a solution does not produce a significant amount of \NV\ and/or \OVI.

In Table~\ref{tab_models} we summarize the PI model parameters. The large total hydrogen column densities (i.e. $\log N_{\rm H} > 19.0$) suggest that a significant amount of baryons is associated with both phases. The thicknesses ($L_{\rm los}$) of the two phases are tens of kpc and differ from each other by only a factor of $\sim3$. But, the density of the low-ionization phase is $\sim 30$ times higher than that of the high-ionization phase. Most interestingly, the metallicity of the high-ionization phase is over an order of magnitude higher than the low-ionization gas phase. The data clearly indicate different origins of the high- and low-ionization gas phases.

%% 
%==================================================================================
%==================================================================================
%%  

%= = = = = = = = = = = = = = = = = = = = = = = = = = = = = = = = = = = 
\begin{table} 
\begin{center}  
\caption{Summary of PI model parameters.}   
\begin{tabular}{cccrcc}  
\hline \hline 
Phase &  $\log n_{\rm H}$ &  $\log U$  &  $>\rm [X/H]$  &  $>\log N_{\rm H}$  &  $>L_{\rm los}$  \\     
      &    (cm$^{-3}$)    &            &         &   (cm$^{-2}$)      &  (kpc)          \\        
\hline 
~\\ 
High  &  $-4.1$          &  $-1.75$   &  0.3    &  19.1              &  53             \\   
Low   &  $-2.6$          &  $-3.25$   & $-1.4$  &  20.2              &  19             \\ 
~\\ 
\hline   
\label{tab_models}  
\end{tabular}        
\end{center}  
\end{table}  
%= = = = = = = = = = = = = = = = = = = = = = = = = = = = = = = = = = = 

\section{Discussion}        
\label{sec_diss}  

\subsection{The strong and large velocity spread {\rm \OVI} absorption}   

We have presented a detailed analysis of an ultra-strong intervening \OVI\ absorber at \zabs\ = 0.39853 detected in the $HST/$COS spectrum of Q~0122--003. The absorber exhibits $\log N(\OVI) = 15.16\pm0.04$ with a kinematic spread of $\Delta v_{90} = 419$~ \kms. Such a strong (i.e. $\log N > 15.0$) and large velocity spread (i.e. $>400$~ \kms) intervening \OVI\ systems are very rare in the Milky Way halo \citep[]{Wakker03,Savage03}, in high-velocity clouds \citep[HVCs;][]{Sembach03}, in the CGM \citep[]{Werk14} and in the IGM \citep[]{Tripp08,Muzahid12a,Savage14}. For example, only $2/42$ systems in the COS-Halos sample show $\log N(\OVI) > 15.0$ \citep[]{Tumlinson11sci}. In a blind survey of intergalactic \OVI\ absorbers at high-$z$ ($1.9 < z < 3.1$) \citet{Muzahid12a} have found only $3/85$ systems with $\Delta v_{90} > 200$~ \kms\ with none of them having  $\Delta v_{90} > 400$~ \kms. The systems with large $\Delta v_{90}$, in their sample, are thought to be associated with outflows from Lyman Break Galaxies \citep[LBGs; see also][]{Adelberger03,Adelberger05}. The only known \OVI\ absorbers at low-$z$ with $\Delta v_{90}$ in excess of 400 \kms\ are the \zabs\ = 0.3558 system towards J~1009+0713 \citep[]{Tumlinson11}, the \zabs\ = 0.927 system towards PG~1206+459 \citep[]{Fox13} and the \zabs\ = 0.227 systems towards the QSO pair Q~0107--025A and Q~0107--025B \citep[]{Muzahid14}. \citet{Tripp11} have argued that a post-starburst outflow is responsible for the wide velocity spread absorption seen towards PG~1206+459 \citep[but see][for an alternate explanation]{Ding03b}. \cite{Tumlinson11} have suggested that \OVI\ arises from interfaces between ``cool" and ``hot" gas phases. Moreover, \cite{Muzahid14} has conjectured that the large velocity spread \OVI\ absorption seen through the pair of lines-of-sight is originating from an ``ancient outflow" \citep[see e.g.][]{Ford14}.

Recently, in 15 of 20 Lyman limit systems (LLS) with $\log N(\HI) > 17.3$ at $2 < z < 3.5$, \cite{Lehner14} have reported presence of strong (e.g. average $\log N(\OVI) = 14.9\pm0.3$) and wide velocity spread i.e. $200 < \Delta v < 400$ \kms\ \OVI\ absorption in the ISM/CGM of high-$z$ galaxies and hypothesized that these strong CGM \OVI\ absorbers probe outflows of star-forming galaxies \citep[see also][]{Fox07b}. We note that, a vast majority of local starburst galaxies show ultra-strong ``down-the-barrel" \OVI\ absorption with an average column density of $\log N(\OVI) = 15.0\pm0.3$ \citep[]{Grimes09}. Additionally, these \OVI\ absorption lines are broad with a FHWM of $\sim 367 \pm 95$ \kms.

In summary, intervening \OVI\ absorbers with $\log N > 15.0$ and $\Delta v_{90} > 400$ \kms, as observed in the present system, are extremely rare. Large-scale galaxy outflows are thought to be the most likely origin of such \OVI\ absorbers whenever detected. The frequent occurrence of strong and large velocity spread \OVI\ absorption in local starburst galaxies further strengthens this idea.

\subsection{The presence of strong {\rm \NV} absorption}  

Unlike most low-$z$ intervening \OVI\ absorbers, the present system shows an ultra-strong ($\log N(\NV)=$~ $14.69\pm0.07$) and wide-spread ($\Delta v_{90} = 285$ \kms) \NV\ absorption with $\log N(\NV)/N(\OVI)$~ $= -0.47 \pm 0.08$. As the creation and destruction ionization potentials of \NV\ are 77.5~eV and 97.9~eV respectively, a hard radiation field at energies above 4~Ryd is required to ionize nitrogen to its {$\rm N^{+4}$} ionization state. Though no systematic survey exists, weak \NV\ absorption ($<10^{14}$ \sqcm) is detected in only a handful of \OVI\ systems at low-$z$ (e.g. \zabs\ = 0.20260 system towards PKS~0312--77 \citep[]{Lehner09}, \zabs\ = 0.16716 system towards PKS~0405--123 \citep[]{Savage10}, \zabs\ = 0.227 towards Q~0107--025A \citep[]{Muzahid14}). In all of them \OVI\ is found to be fairly strong ($>10^{14.7}$ \sqcm). Here, we note that \cite{Werk13} have not found any \NV\ absorption with $\log N >14$ in the COS-Halos sample.

To our knowledge, there are only 6 other intervening systems that show $\log N(\NV) > 14.0$ and only one of them is at $z<1$. These are the \zabs\ = 0.927 system towards PG~1206+459 \citep{Tripp11}, \zabs\ = 2.811 system towards Q~0528--250 \citep{Fox07b}, \zabs\ = 2.83437 system towards J~1343+5721 and \zabs\ = 2.47958 system towards Q~1603+3820 and \zabs\ = 2.18076 system towards Q~1217+499 \citep{Lehner14}, and the \zabs\ = 1.5965 system towards PKS~0237--23 \citep{Fechner09}. Interestingly, all these systems show strong \OVI\ ($\log N > 14.8$) absorption kinematically spread over $>200$~ \kms\ similar to the present system\footnote{Note that \OVI\ information is not available for the \zabs\ = 1.5965 system towards PKS~0237--23. However it shows a strong \CIV\ absorption spread over $\sim300$ \kms. The strong correlation between the line spreads of \CIV\ and \OVI\ absorption found by \cite{Muzahid12a} suggests that this system should have a strong and wide-spread \OVI\ absorption.}, and all of them are LLS.

A population of photoionized intervening \NV\ absorbers, with $d{\mathcal N}/dz\sim 0.9$ down to $\log N = 12.7$, at high-$z$ ($1.5 < z < 2.5$) is reported by \citet{Fechner09}. These absorbers are predominantly weak and only one out of 21 systems shows $\log N(\NV) > 14.0$. As the intensity of the extra-galactic UV background radiation is considerably higher as compared to low-$z$ due to the enhanced QSO activity at $z\sim2$ \citep[]{Haardt96,Khaire15}, detection of weak \NV\ absorption at high-$z$ is consistent with expectations for PI models. However, the presence of \NV\ with $\log N > 14.0$ is very rare at any redshift.

Apart from the present system, the only other \OVI\ absorber with $\log N(\NV) > 14.0$ at $z < 1.0$ is the well known system at \zabs\ = 0.927 towards PG~1206+459 \citep{Churchill99b,Ding03b,Tripp11}. These two systems have remarkable resemblance in the absorption properties of their high-ions and show $\log N(\NV)/N(\OVI) \sim -0.5$ (Rosenwasser et al., in preparation). Unlike \cite{Tripp11}, who favored a collisional ionization scenario for \NV\ and could not estimate a high-ionization phase metallicity, Rosenwasser et al. have found that both \OVI\ and \NV\ can be explained under PI equilibrium with solar to super-solar metallicities and with a density of $\sim 10^{-4}$ cm$^{-3}$. As we discuss next, all these properties are strikingly similar to what we deduce for the present system.

Very strong \NV\ absorption is characteristic of many intrinsic narrow and mini-broad absorption line systems, that are known to be related to the AGN winds/host-galaxy because of a velocity within $\sim$~5000 \kms\ of the QSO redshift, and evidence of partial coverage of the QSO continuum/broad emission line source \citep[]{Hamann00,Srianand02,Misawa07,Wu10}. For example, \cite{Wu10} reported three $2.6 < z < 3.0$ intrinsic \NV\ absorbers with metallicities greater than 10 times the solar value. Strong \OVI\ and \CIV\ absorption are common in these systems as well. The present system is clearly not an intrinsic absorber, but the common properties are likely evidence of an origin in the high metallicity, central region of a galaxy. Both the material that feeds an AGN wind, and that which is ejected in a starburst outflow share a similar chemical origin in a high metallicity environment \citep[]{Hamann99}.

\subsection{Origins of different gas phases}             

In Section~\ref{sec_avgmodels} we have demonstrated that the \OVI\ bearing gas in this absorber can be explained with PI equilibrium models that require a density of $\sim 10^{-4.1}$~cm$^{-3}$, a super-solar metallicity (i.e. $\rm [X/H] \gtrsim 0.3$), and a solar $\rm [N/O]$ ratio. The synthesis of N can arise from primary and secondary production via the CNO cycle. Primary production is synthesis of N from C produced in the core of the same star via helium burning and secondary production is the synthesis of N from C and O produced in previous generations of stars. Secondary N enrichment of the ISM occurs well after massive stars have gone Type-II supernovae and seeded the ISM with oxygen. For enrichment from primary synthesis, the $\rm [N/O]$ ratio is roughly $-0.6$ for $\rm [O/H] \lesssim -0.3$. For enrichment from secondary synthesis, the $\rm [N/O]$ ratio increases with increasing metallicity for $\rm [O/H] \gtrsim -0.3$, such that by $\rm [O/H] \simeq +0.3$, $\rm [N/O] \simeq 0.0$ i.e. the solar value \citep[e.g. see Figure~9 of][]{Pettini08a}. Therefore, the assumption of a solar $\rm [N/O]$ ratio in this system is reasonable. Solar/super-solar metallicity with solar $\rm [N/O]$ ratio suggests that nitrogen in the high-ionization absorbing gas is predominantly produced via the secondary channel. Thus, the high-ionization, low density gas phase is likely to be originating from a region of the galaxy with a prolonged and high star formation rate \citep[see also][]{Hussain15}. This is consistent with the host-galaxy properties as discussed in the next section.

It is often claimed that strong \OVI\ absorbers are collisionally ionized \citep[e.g.][]{Lehner14} as they require unreasonably large sizes under PI equilibrium \citep[but see][]{Muzahid14}. However, from the observed narrowness of the \lyb\ profile, the non-detection of \SIV\ absorption, and the presence of strong \CIV\ absorption in the low-resolution FOS spectrum, we have ruled out the possibility of \OVI\ bearing gas in this system being collisionally ionized (Appendix~\ref{app_CIE} and \ref{app_nCIE}).

For the low-ionization phase, both the overall (Section~\ref{sec_avgmodels}) and the component-by-component (Appendix~\ref{app_photlow}) approach of PI modeling suggest a metallicity of $\rm [X/H] \gtrsim -2.0$~to~$-1.3$. The high-ionization phase has more than a factor of $\sim10$ higher metallicity. This is in contrast to the \zabs\ = 0.927 system towards PG~1206+459 for which Rosenwasser et al. (in preparation) have found that the solar/super-solar metallicity of the low-ionization gas is indeed similar to the high-ionization gas. We note that in the present case there is no one-to-one correspondence between the high- (e.g. \OVI) and low- (e.g. \MgII) ionization absorption kinematics. But in the case of PG~1206+459 kinematic similarity between \MgII\ and \OVI\ absorption is quite remarkable. Therefore, while for PG~1206+459 both the high- and low-ionization gas phases are consistent with being part of same multiphase outflow \citep[e.g.][]{Tripp11}, the low-ionization phase in the present system likely has a very different origin from the high-ionization phase.

Using the methods of \cite{Kacprzak10a}, we measured the rotation curve of the host-galaxy from the H$\alpha$ emission line, which extends $\pm 1''$ (corresponding to 3.5 kpc) from the galaxy center. The projected rotation velocity is $\simeq 200$~\kms. We then employed the rotation model of \cite{Steidel02} to examine the range of allowed velocities along the line-of-sight that are consistent with extended galaxy rotation given the galaxy inclination, azimuthal angle, the quasar impact parameter, and the gas scale height, $h_\nu$. Following \cite{Kacprzak10a}, we use $h_\nu=1$ Mpc (for a non lagging halo above the galaxy plane). The range of velocities along the line-of-sight that could be due to extended rotation from the galaxy are $+25 \leq v \leq +150$~{\kms}. The components of the low-ionization gas are at $v ~\simeq 0$ and $v \simeq 200$~ {\kms}, as probed with {\MgII} absorption. None of the \MgII\ components is consistent with co-rotation of the galaxy. The low metallicity and kinematics of this phase perhaps suggest that it stems from recycled gas as seen in numerical simulation of \cite{Oppenheimer10}. Such an interpretation is also supported by the recent study of ``bimodal" metallicity distribution of low-ions in low-$z$ LLS by \cite{Lehner13}. The metal-poor branch, possibly tracing cold accretion streams, in their sample peaks at $\rm [X/H] = -1.6$ which is consistent with what we find for this system.

\subsection{The host-galaxy}   

At the epoch of our measurement, the host galaxy has a high SFR of $6.9~M_\odot$ yr$^{-1}$ and a $\Sigma_{\ast}$ of $\sim0.4~M_{\odot}$ yr$^{-1}$ kpc$^{-2}$, the latter being a factor of $\sim 4$ above the well known threshold of 0.1 $M_{\odot}$ yr$^{-1}$ kpc$^{-2}$ for driving galactic-scale winds \citep[]{Heckman03}. The observed galaxy orientation parameters, i.e. $i = 63\degree$ and $\Phi = 73\degree$, as given in Table~\ref{tab:zfield} suggest that the QSO line-of-sight is at ideal position to pierce one of the bicones of a bi-conical outflow propagating along the minor-axis of the galaxy \citep[e.g.][]{Bouche12,Kacprzak12a,Kacprzak14,Bordoloi14a,Schroetter15}. The host-galaxy has a metallicity of $\rm 12+ \log (O/H) =$~ $8.68\pm0.02$, i.e. $\rm [O/H] \sim 0.0$, which is roughly consistent with the metallicity of high-ionization gas phase. All these observational evidences strongly indicate that the high-ionization gas, as probed by the ultra-strong \OVI\ and \NV\ absorption, is tracing a powerful (see the next section), active, metal-enriched, large-scale galactic outflow.

Here we emphasize that the absorber is unlikely from the ISM of a dwarf galaxy that happens to be coincident on the quasar. \cite{vanZee06} found that the H~{\sc ii} regions of dwarf galaxies have sub-solar metallicity and $\rm [N/O]$ ratios consistent with primary N synthesis.  On the other hand, \cite{vanZee98} showed that the H~{\sc ii} regions of massive spiral galaxies have solar to super-solar metallicity and $\rm [N/O]$ ratios consistent with secondary N synthesis.  As such, we favor the scenario in which the high-ionization absorbing gas was ejected from the host spiral galaxy we have identified at the absorption redshift.

\subsection{The outflow properties}  

From the halo mass ($M_{h} \sim 10^{12.5} M_{\odot}$) and the virial radius ($R_{\rm vir} \sim 280$~ kpc) of the host-galaxy we estimate a circular velocity of $v_{c} \sim 220$~ \kms. Using the relationship between SFR, $v_c$, and the terminal outflow speed ($v_w$) by \citet{Sharma12} we estimate $v_w \sim 230$ \kms. Here we note that $v_c$ is calculated at $R_{\rm 200}$ in their models whereas we compute $v_c$ at the virial radius. Nevertheless, this does not change our estimated $v_w$ by more that a few percent. An outflow with a constant speed of 230~ \kms\ would require 0.7 Gyr time to travel the minimum projected distance of 163 kpc. This is on the order of the sound crossing time for the individual cool ($T\sim10^{4}$~K) photoionized clouds with size $\sim 4-10$~ kpc (see Appendix~\ref{app_photlow}). However, in practice the outflow could have been ejected with a significantly higher speed as compared to the terminal speed we estimate here. In that case the flow time could be significantly lower as compared to the sound crossing time so that the clouds would have enough time to be stable against any mechanical disturbances before they traverse a distance of 163~kpc.

Under the thin-shell approximation the mass of the outflowing gas can be written as: $M_{\rm out} = 4 \pi \mu m_p C_{\Omega} C_{f} N_{\rm H} r^{2}$, where $m_p$ is the proton mass, $\mu = 1.4$ accounts for the mass of helium, $C_{\Omega}$ $(C_{f})$ is the global (local) covering factor, and $r$ is the thin-shell radius \citep[e.g.][]{Rupke05}. The global covering factor is related to the wind's opening angle. Here we assume $C_{\Omega} = 0.4$ as observed in local starbursts \citep[e.g.][]{Veilleux05}. The local covering factor, on the other hand, is related to the wind's clumpiness. For simplicity we assume it to be unity for the high-ionization, low density, diffuse gas. Using $r=$~163 kpc and $N_{\rm H} = 10^{19.1}$~ \sqcm\ (see Table~\ref{tab_models}) we estimate $M_{\rm out} \sim 2\times10^{10} M_{\odot}$ corresponding to an oxygen mass of $M_{\rm O} \sim 10^{7}~ M_{\odot}$. These are consistent with the masses as estimated in the CGM of $\sim L_{\ast}$ galaxies at $z \lesssim 0.2$ by \cite{Tumlinson11sci}. Using the wind speed of 230 \kms, we further estimate the mass-flow rate of $\dot{M}_{\rm out} \sim 54~ M_{\odot}$~ yr$^{-1}$ and the kinetic luminosity of $\dot{E_{k}} \sim 9\times10^{41}$ erg~s$^{-1}$ (see Appendix~\ref{app_equations}). These values are typical of what have been seen in ``down-the-barrel" outflows from infrared-luminous starbursts at $z < 0.5$ \citep[i.e.][]{Rupke05}.

To better understand how the mass outflow rate is related to the SFR of the host galaxy, it is customary to define the mass loading factor, $\eta \equiv \dot{M}_{\rm out}/\rm SFR$, which is a critical ingredient for numerical simulations of structure formation \citep[e.g.][]{Oppenheimer10,Dave11b}. We estimate a $\eta$ of $\sim 8$ for the outflow studied here. In the infrared-luminous starbursts sample of \cite{Rupke05}, $\eta$ is found to range from 0.001 to 10 but is $\sim0.1$ on an average. However, we note that a vast majority of these starburst galaxies have SFR $> 100~M_{\odot}~ \rm yr^{-1}$ which is significantly higher as compared to the present galaxy with SFR of $6.9 ~ M_{\odot} ~ \rm yr^{-1}$. Therefore, the galaxy in our study is extremely efficient in entraining mass, in the form of outflow, for its SFR. This probably because of the high $\Sigma_{\ast}$. However, it might also be possible that the host-galaxy had much higher SFR at the epoch of ejection compared to what we observe today.                                                
   
\section{Summary} 
\label{sec_summ}   

We have examined the physical conditions, chemical abundances, and energetics of a large-scale galactic outflow in the CGM of a star-forming (SFR~$6.9~M_{\odot}$~yr$^{-1}$), sub-$L_{\ast}$ ($0.5 L_{B}^{\ast}$) galaxy at $z = 0.39853$. Using the halo abundance matching technique we estimate a halo mass of $M_{h} \sim 10^{12.5 \pm 0.2} M_{\odot}$ and a virial radius of $R_{\rm vir} \sim 280$ kpc for the galaxy. Along with \lya\ and \lyb, several low- (i.e. \MgI, \MgII, \CII, \NII, \SiII, \SiIII) and high- (i.e. \CIV, \NV, \OVI) ionization metal absorption lines, originating from the CGM of the galaxy, have been detected in the $HST$(COS, FOS) and $VLT$(UVES) spectra of the background quasar Q~0122--003. The QSO is located at an impact parameter of $D = 163$ kpc ($D/R_{\rm vir} = 0.6$) with an azimuthal angle of 73$\degree$ with respect to the galaxy major-axis. Our main findings are as follows:

\begin{enumerate}
 
\item The low-ionization absorption components trace photoionized gas with a density of $n_{\rm H} \sim 10^{-2.6}$~cm$^{-3}$ ($\log U \sim -3.2$) and a metallicity of $\rm [X/H] \gtrsim -1.4$. Such a low metallicity gas is consistent with being recycled material in the galaxy halo as predicted in numerical simulations. 

\item The high-ionization gas phase shows ultra-strong \OVI\ with $\log N(\OVI) = 15.16\pm0.04$ kinematically spread over $\Delta v_{90} = 419$~ \kms. Such a strong and large velocity spread \OVI\ absorber is very rare at any redshift. Among the known astrophysical environments in the local universe, only starburst galaxies show such strong \OVI\ absorption \citep[]{Grimes09}.           

\item Unlike most intervening \OVI\ systems, this absorber shows ultra-strong and wide velocity spread \NV\ absorption with $\log N(\NV) = 14.69\pm0.07$ and $\Delta v_{90} = 285$~ \kms. There is only one system at $z < 1$ \citep[i.e. \zabs\ = 0.927 towards PG~1206+459,][]{Tripp11} that shows \NV\ as strong as this.   

\item It is often claimed that strong \OVI\ absorbers are collisionally ionized as they commonly require large sizes ($>$ Mpc) under PI equilibrium \citep[]{Lehner14}. However, for this system our PI models suggest line-of-sight thicknesses for individual components of $\sim10$~kpc. In fact, based on the $b$-parameter constraints from the \lyb\ profile, the non-detection of \SIV, and the presence of strong \CIV\ ($W_{r} = 1.7$ \AA) in the FOS spectrum, we have ruled out CIE and non-CIE models. 

\item The high-ions, \OVI\ and \NV, can be well explained as arising in a low-density ($n_{\rm H} \sim 10^{-4.2}$~cm$^{-3}$, $\log U \sim -1.6$) photoionized gas with super-solar metallicity ($\rm [X/H] \gtrsim 0.3$) and with solar $\rm [N/O]$ ratio. The super-solar metallicity with a solar $\rm [N/O]$ ratio implies that nitrogen in this system is predominantly produced via secondary synthesis. Thus the high-ionization, low-density gas is presumably stemming from regions of the host-galaxy that have sustained a high star-formation rate for a prolonged period.   

\item The measured $\Sigma_{\ast}$ of $\sim 0.4$ $M_{\odot}~\rm yr^{-1}~ \rm kpc^{-2}$ for the host galaxy is a factor of $\sim4$ higher than the threshold of $\sim 0.1~ M_{\odot}~ \rm yr^{-1}~ kpc^{-2}$ for driving galactic-scale wind as seen in local starbursts. The azimuthal angle of $73\degree$ suggests that the QSO line-of-sight is in ideal position to pierce the bi-conical outflow propagating along the minor-axis of the host-galaxy.  

\item Assuming the high-ionization gas is in outflow, we estimate a outflow mass of $M_{\rm out} \sim 2\times10^{10}~ M_{\odot}$, mass outflow rate of $\dot{M}_{\rm out} \sim 54~M_{\odot} \rm yr^{-1}$, and kinetic luminosity of $\dot{E}_{k} \sim 9\times10^{41}$ erg~s$^{-1}$ using a thin-shell model. These values are consistent with that of ``down-the-barrel" outflows from infrared-luminous starbursts at $z < 0.5$ \citep[]{Rupke05}.   

\item We estimate a mass loading factor of $\eta \sim 8$. This is among the highest values as seen in infrared-luminous starbursts at low-$z$. However, we note that the SFR of the host-galaxy is significantly lower (by a factor of ten or more) than the starburst galaxies \citep{Rupke05}. This indicates that the host-galaxy is highly efficient in entraining mass onto the CGM in the form of outflow.    

\end{enumerate}  

\vskip-0.2cm 
Finally, we emphasize that such powerful, large-scale, metal-rich outflows are the essential component by means of which a galaxy imparts sufficient mechanical and chemical feedbacks that regulates star formation and hence evolution of a galaxy. Finding and analyzing more such unique absorption systems is crucial for a comprehensive understanding of galaxy feedback mechanisms and reinforce galaxy evolution theory with useful observational constraints.

\vskip0.4cm 
Support for this research was provided by NASA through grants HST GO-13398 from the Space Telescope Science Institute, which is operated by the Association of Universities for Research in Astronomy, Inc., under NASA contract NAS5-26555. GGK acknowledges the support of the Australian Research Council through the award of a Future Fellowship (FT140100933). Some of the data presented here were obtained at the W. M. Keck Observatory, which is operated as a scientific partnership among the California Institute of Technology, the University of California and the National Aeronautics and Space Administration. The Observatory was made possible by the generous financial support of the W. M. Keck Foundation. Observations were supported by Swinburne Keck programs 2014A\_W178E, 2014B\_W018E, and 2015A\_W018E.

\vskip0.2cm 
\noindent 
{\it Facilities:}~$\it HST$(COS,~WFPC2),~$\it KECK$(ESI),~$\it VLT$(UVES)

\vskip-0.2cm 
\appendix

\section{Estimation of $N(\HI)$ in different low- and high-ionization components}   
\label{app_NHIestimate}

%% 
%==================================================================================
\begin{figure*} 
\centerline{\vbox{
\centerline{\hbox{ 
\includegraphics[width=0.33\textwidth,angle=00]{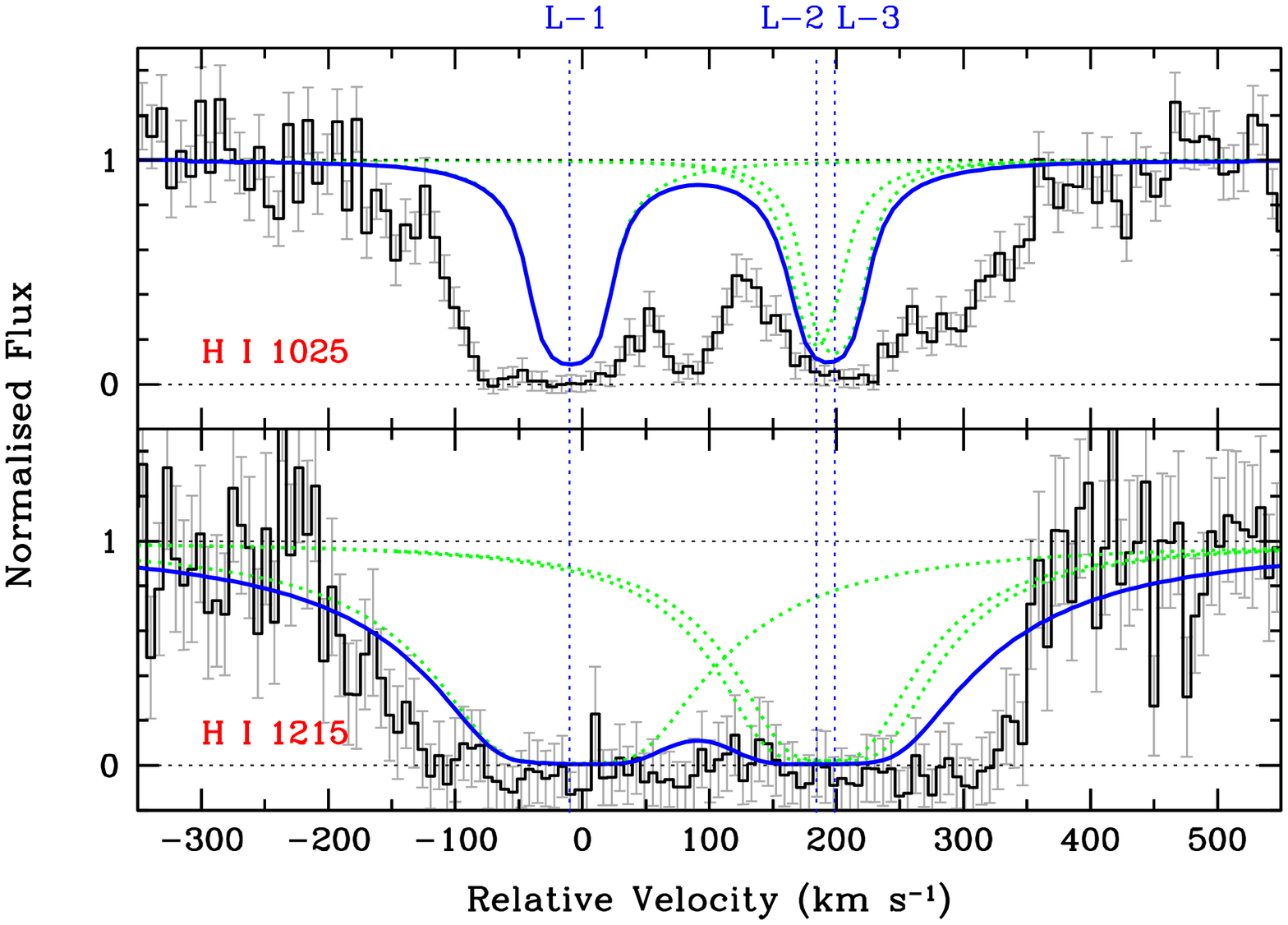} 
\includegraphics[width=0.33\textwidth,angle=00]{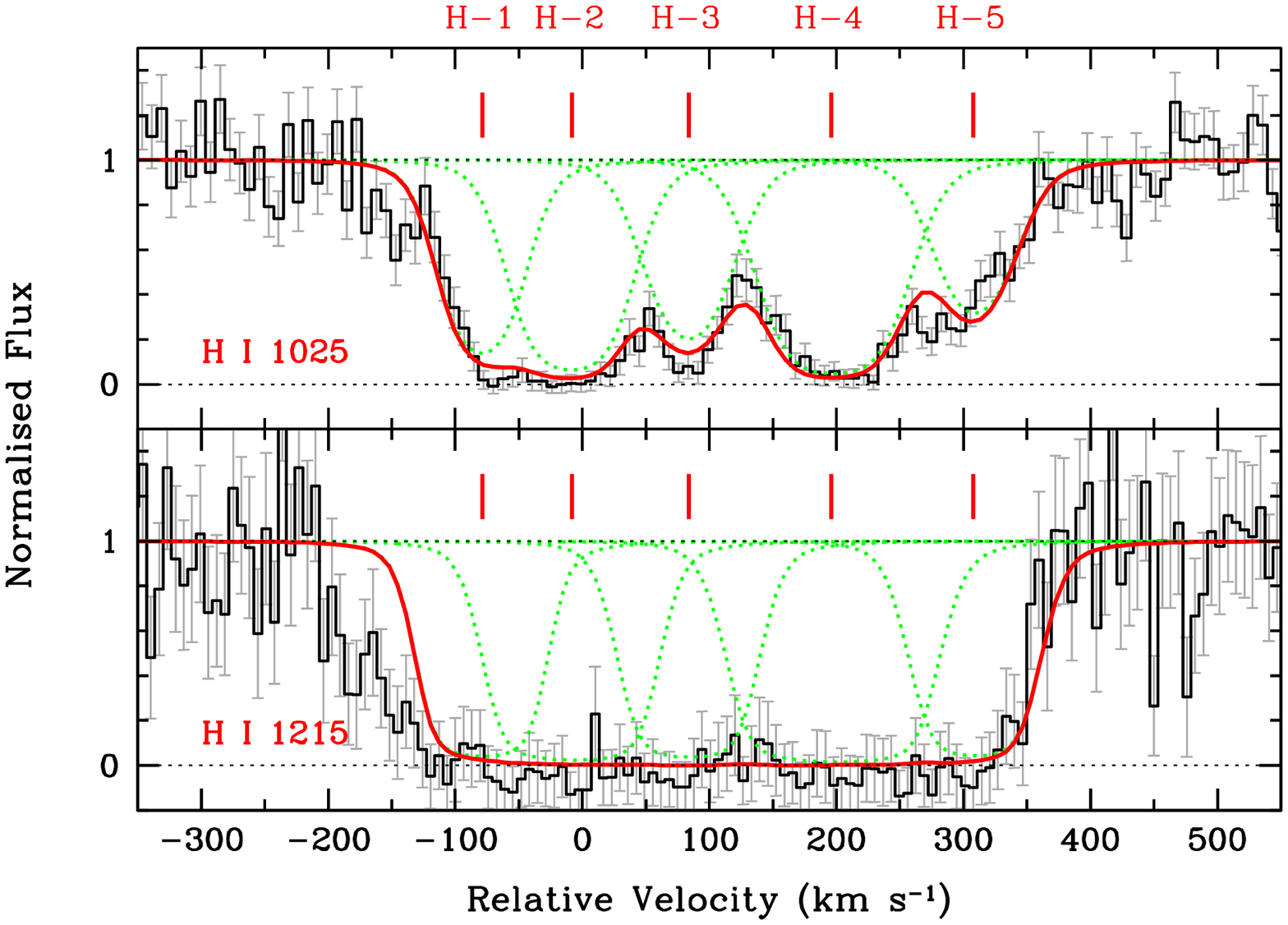} 
\includegraphics[width=0.33\textwidth,angle=00]{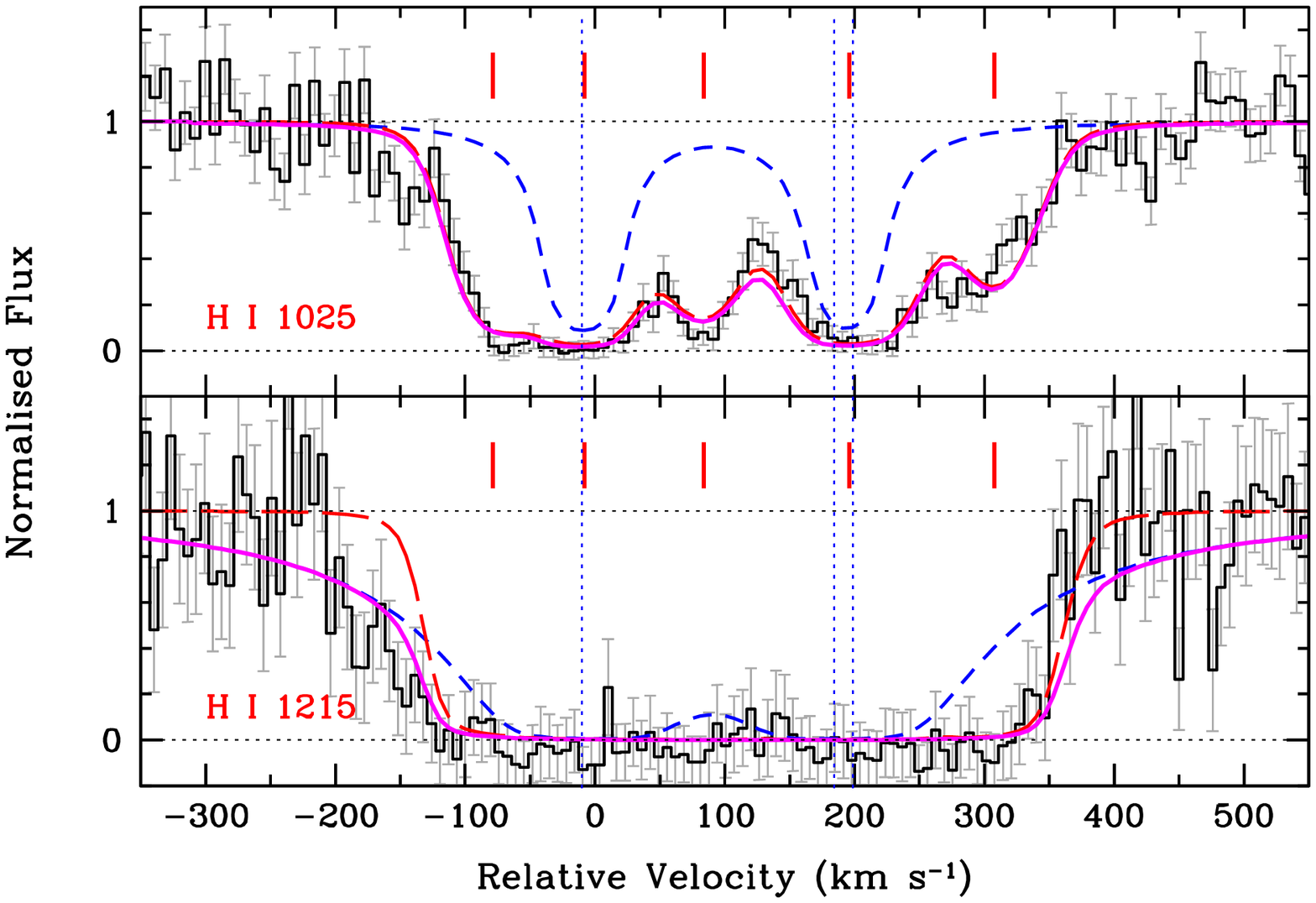}
}}
}}   
\caption{Estimation of upper limits on $N(\HI)$ in different low- (left panel) and high- (middle panel) 
ionization absorption components. Individual profiles are shown in (green) dotted curves. The resultant 
profiles are shown in solid curves. The error in each pixel is shown as a grey error-bar. The vertical dotted 
lines in the left panel are line centroids of the \MgII\ components. The small ticks in the middle panel 
are line centroids of the \OVI\ components. Both low- and high-ionization components are shown in the 
right panel. Contributions from low- and high-ionization components are shown in (blue) dotted and 
(red) dashed curves respectively. The resultant solid (magenta) curve explains both the \lya\ and \lyb\ 
profiles satisfactorily.          
}      
\label{HI} 
\end{figure*} 
%==================================================================================
%%  

In this section we describe how $N(\HI)$ values in different low- and high-ionization absorption components are estimated. Determining $N(\HI)$ component-by-component in a multiphase, multicomponent absorption system is extremely challenging. The presence of unsaturated higher-order Lyman series lines is very helpful in order to disentangle the contributions from high- and low-ionization phases \citep[see e.g.][]{Muzahid14}. Our present COS spectrum only covers \lya\ and \lyb\ absorption for this absorber. We, therefore, try to estimate conservative upper limits on $N(\HI)$ in different low- (L--1 to L--3) and high- (H--1 to H--5) ionization components.

To estimate the maximum allowed $N(\HI)$ associated with the low-ionization components, we use the \MgII\ component structure and assume pure non-thermal line broadening, i.e. $b(\HI) = b(\MgII)$\footnote{$N(\HI)$ values can be significantly lower if one assumes pure thermal broadening, i.e. $b(\HI) = \sqrt{24}~ b(\MgII)$. However, the wings of the \lya\ absorption cannot be explained in such a scenario.}. This gives the lowest possible $b$-parameters for \HI\ in each component. We then use {\sc vpfit} to generate synthetic profiles corresponding to different $N(\HI)$ values. We start with an initial value of $\log N(\HI) = 14.0$ and gradually increase the $N(\HI)$ value until the synthetic profile just exceeds the observed \lya\ and/or \lyb\ profiles. The use of the lowest possible $b(\HI)$ values ensures that the $N(\HI)$ values we report in Table~\ref{tab_lowion} are indeed conservative. The left panel of Figure~\ref{HI} shows the synthetic profiles corresponding to the maximum allowed $N(\HI)$ values we estimate for the low-ionization components. It is apparent from the bottom panel that the wings of the \lya\ absorption are reproduced adequately. Note that because of the assumed low $b$-parameters, $N(\HI)$ value as high as $\sim 10^{18.6}$ \sqcm\ does not produce any significant \lyb\ optical depth. Therefore, almost all the \lyb\ absorption and the remaining absorption seen in \lya\ must arise from the high-ionization components. As mentioned in the footnote of Table~\ref{tab_lowion} our $N(\HI)$ limits remain unaltered for $b$-values in the range 14 -- 20 \kms\ which is expected from the PI model temperatures.

To estimate the maximum allowed $N(\HI)$ corresponding to the high-ionization components, we use the \OVI\ component structure and again assume pure non-thermal line broadening, i.e. $b(\HI) = b(\OVI)$. Similar to the low-ionization components, the assumption of pure non-thermal broadening ensures that the $N(\HI)$ upper limits for the high-ionization components are conservative as well. We fit the remaining absorption (i.e. after accounting for the contribution from the low-ionization components) using {\sc vpfit} by allowing the code to vary $N(\HI)$. The maximum allowed $N(\HI)$ for the high-ionization components are given in Table~\ref{tab_highion}. In the middle panel of Figure~\ref{HI} we show the corresponding profiles as dotted curves. The solid curve represents the resultant profile. We emphasize here that due to the presence of low-ionization components at similar velocities, the $N(\HI)$ limits for the components H--2 and H--4 are uncertain. However, the $N(\HI)$ limits in the rest of the high-ionization components are robust. For the H--1 and H--5 components, the $N(\HI)$ values are constrained adequately by the blue and red wings of the \lyb\ profile, respectively. Moreover, the non-black absorption seen in the \lyb\ at $v \sim +100$~\kms\ provides a robust constraint on the $N(\HI)$ for the H--3 component. A free fit to this part of the \lyb\ absorption, without considering any contributions it might have from the adjacent absorption, leads to a solution with $b(\HI) = 28\pm4$ \kms\ and $\log N(\HI) = 15.10\pm0.10$. This free fit $b$-parameter is consistent with our assumption that $b(\HI) = b(\OVI)$ within $1\sigma$ allowed uncertainty. Note that no low-ions are detected in any of these components (i.e. in \linebreak H--1, H--3, and H--5). Therefore, the observed \HI\ absorption is entirely due to the high-ionization phase. Next, we assume that the \lyb\ profile is completely due to the high-ionization absorption components with $b(\HI) = b(\OVI)$. Under such an assumption the column densities we obtain are very similar to the upper limits we derived in the presence of low-ionization components (see Table~\ref{tab_highion}). In the right panel of Figure~\ref{HI} we show the resultant (high- and low-ionization) profiles which explain both the \lya\ and \lyb\ profiles adequately.

%% 
%==================================================================================
\begin{figure*} 
\centerline{\vbox{
\centerline{\hbox{ 
\includegraphics[width=0.33\textwidth,angle=00]{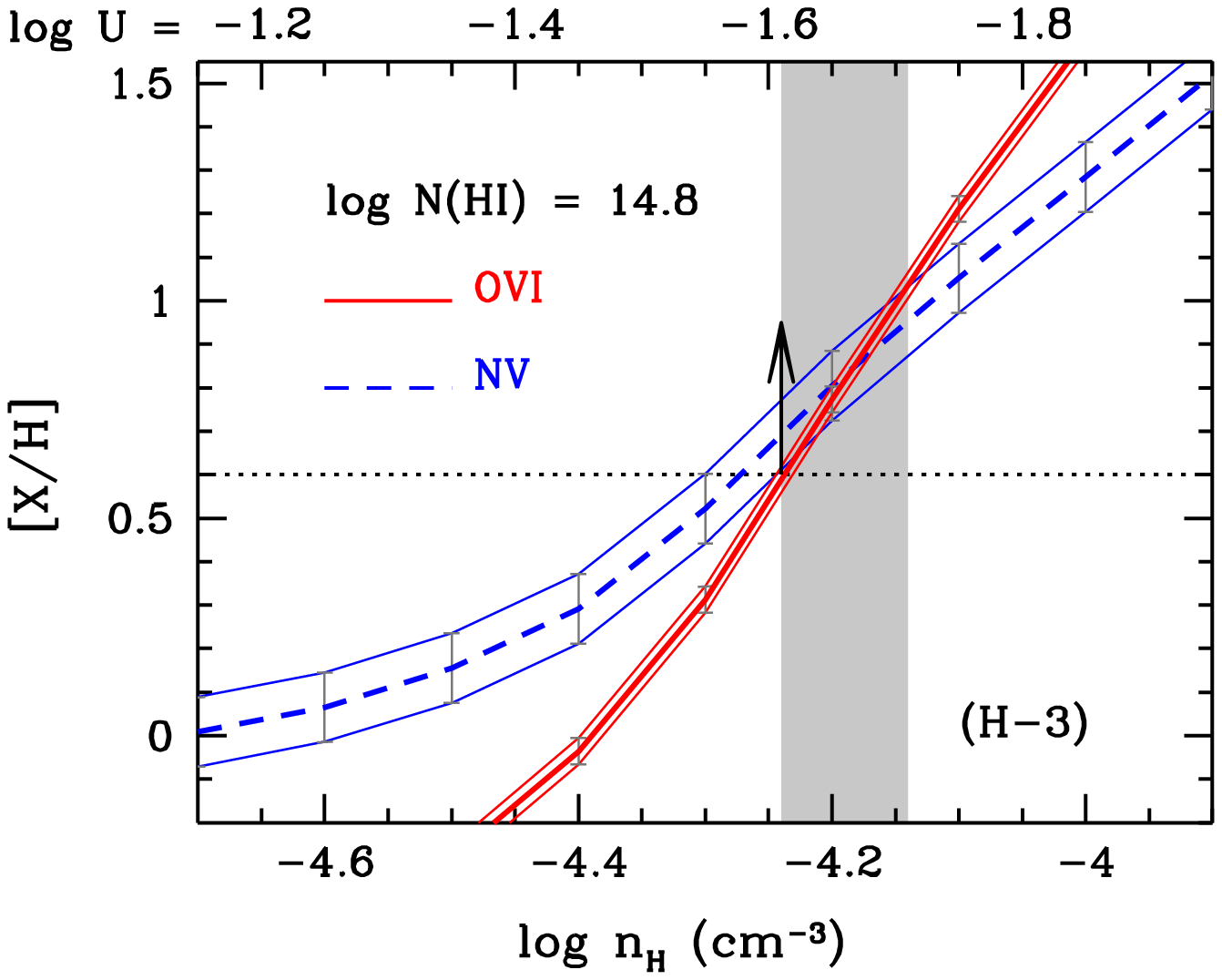}
\includegraphics[width=0.33\textwidth,angle=00]{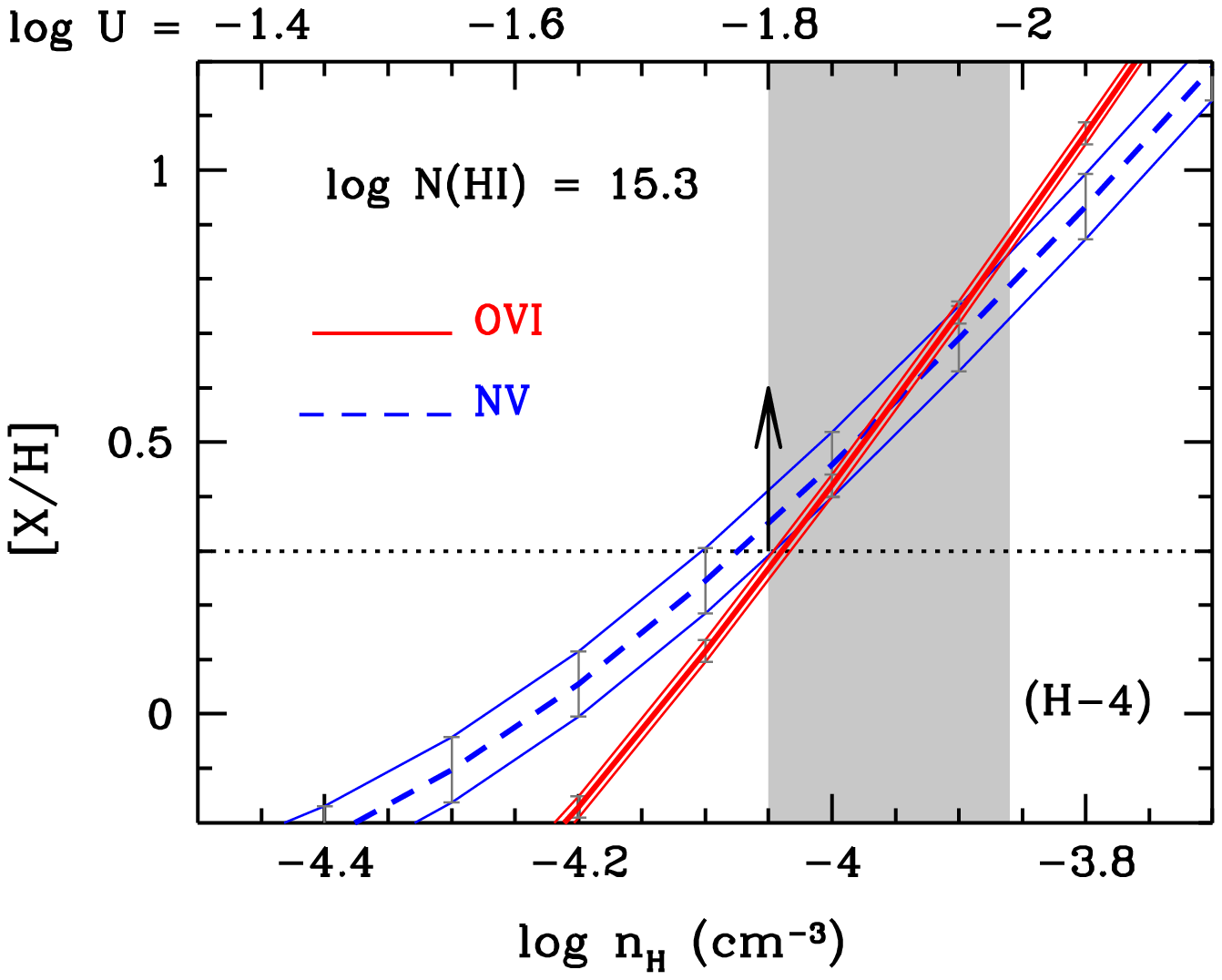}  
\includegraphics[width=0.33\textwidth,angle=00]{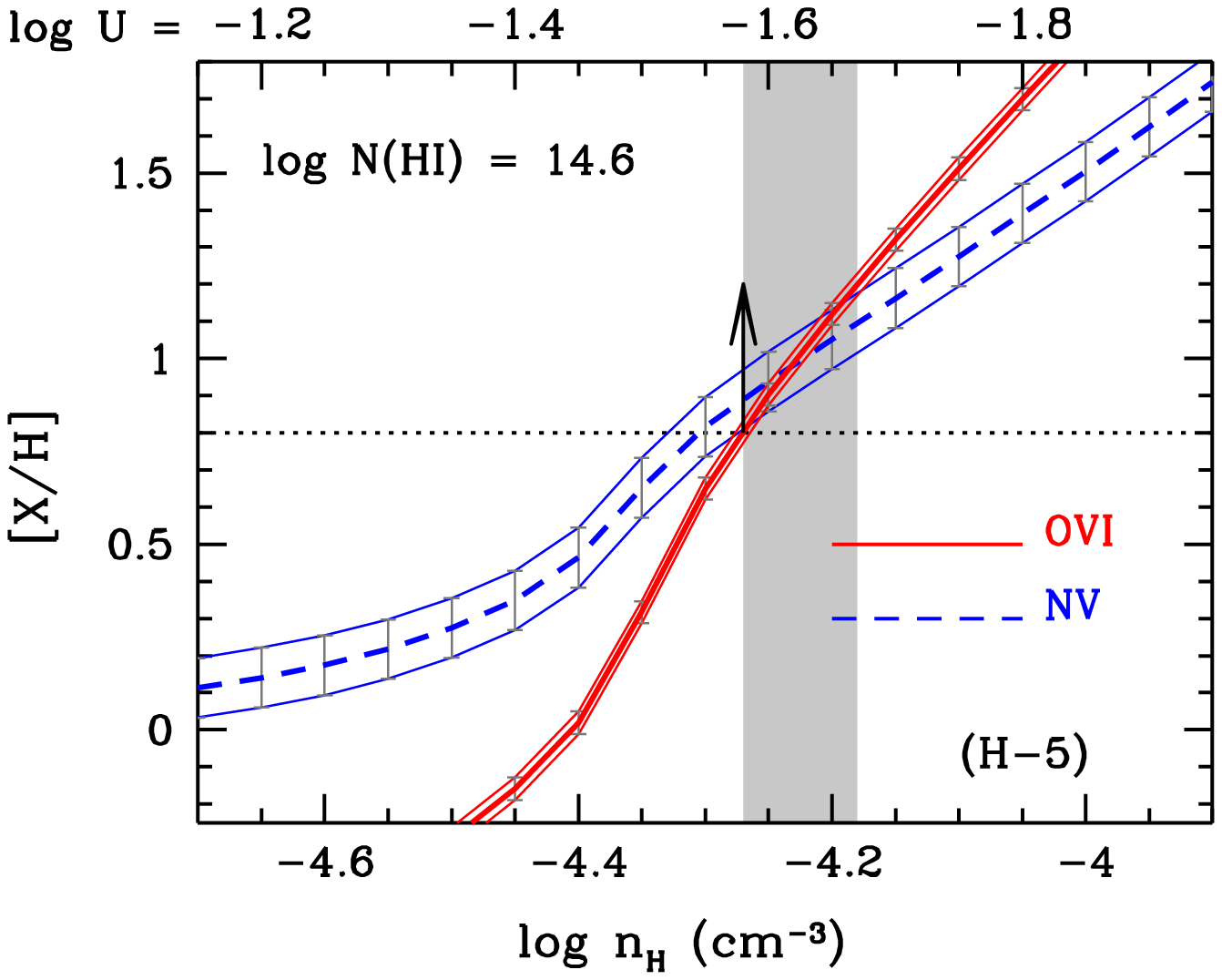} 
}}
}}   
\vskip-0.2cm  
\caption{PI models for the high-ionization absorption components, H--3 (left panel), 
H--4 (middle panel), and H--5 (right panel). Metallicity required to reproduce the observed column 
densities of individual high-ions (\OVI\ and \NV) are plotted as a function of gas density. 
The $N(\HI)$ values assumed by the models are mentioned in the plot (see Table~\ref{tab_highion}). 
The ionization parameters ($\log U$) corresponding to different $n_{\rm H}$ values are labelled at  
the top of each panel. The shaded regions represent the range in $\log n_{\rm H}$ (or $\log U$) 
over which $N(\OVI)$ and $N(\NV)$ match with our measurements. The horizontal dotted line  
represents the lower limit in metallicity indicated by the arrow.}        
\label{Hphot} 
\end{figure*} 
%==================================================================================
%%  

%% 
%==================================================================================
\begin{figure*} 
\centerline{\vbox{
\centerline{\hbox{ 
\includegraphics[width=0.33\textwidth,angle=00]{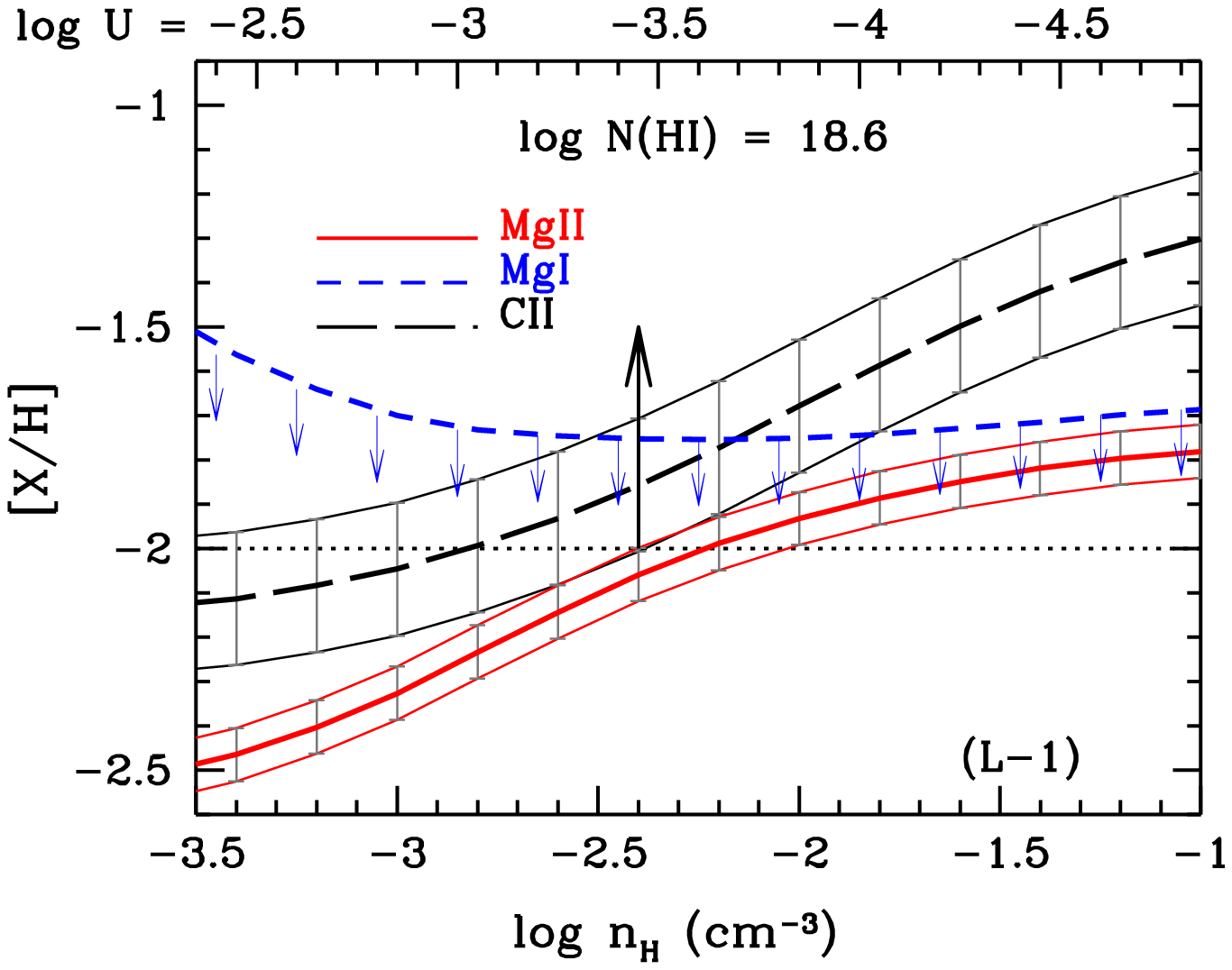}
\includegraphics[width=0.33\textwidth,angle=00]{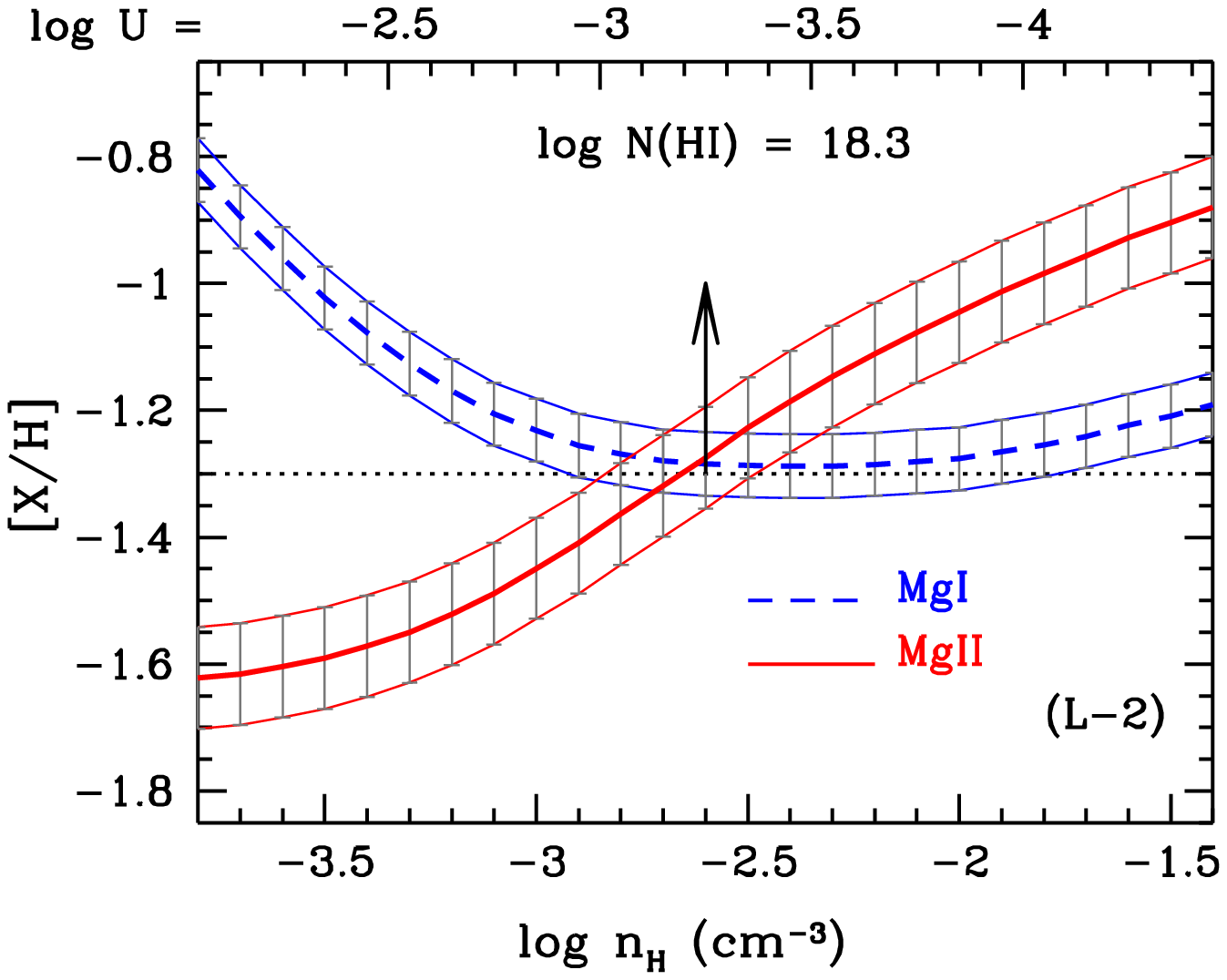}  
\includegraphics[width=0.33\textwidth,angle=00]{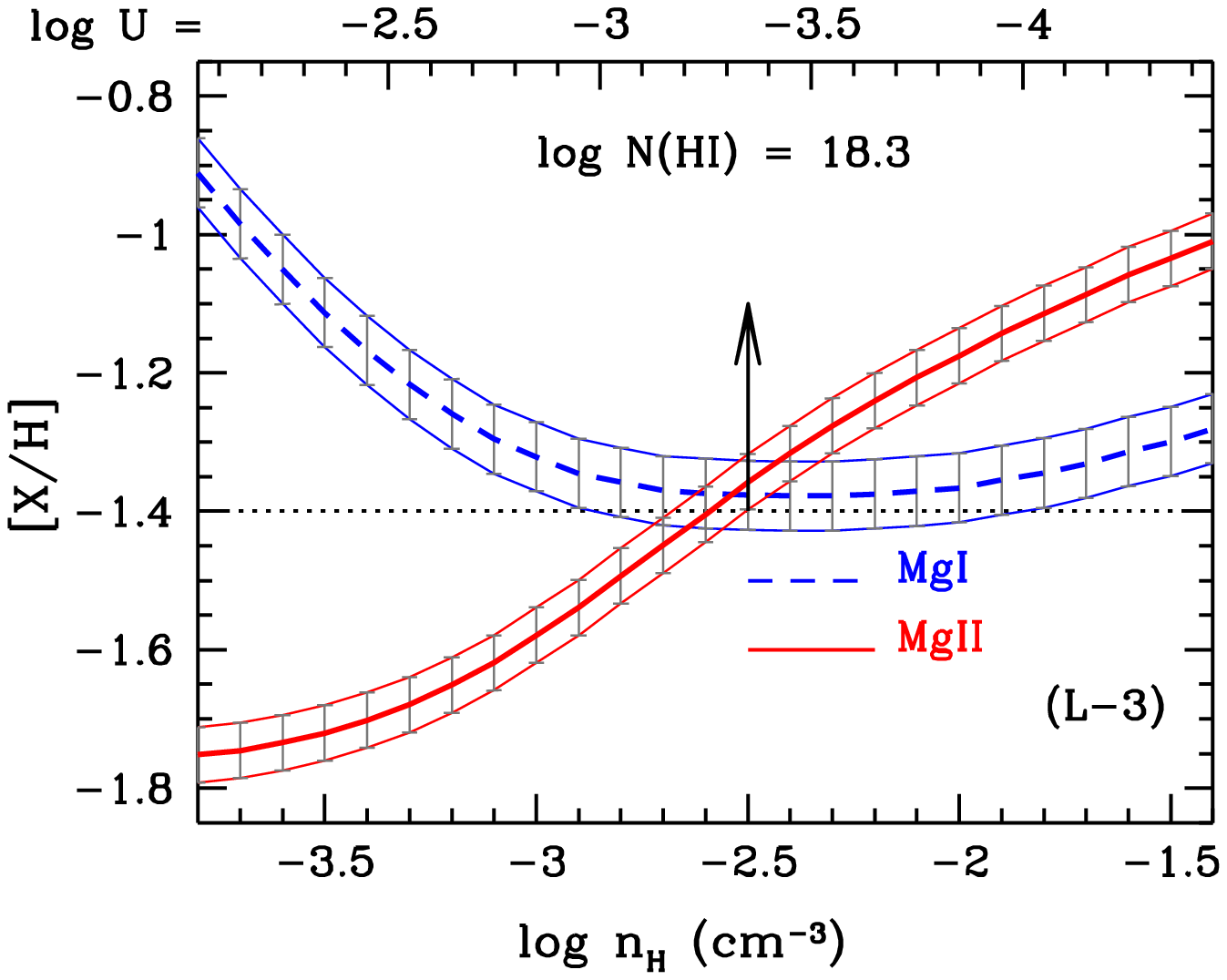} 
}}
}}   
\vskip-0.2cm  
\caption{Same as Figure~\ref{Hphot} but for the low-ionization absorption components, L--1 (left panel),  
L--2 (middle panel), and L--3 (right panel). The measured $N(\MgII)$ and $N(\MgI)$ values along with the 
$N(\HI)$ limits are used to constrain the model parameters. As \MgI\ is not detected in L--1, $N(\CII)$ 
is used instead.}           
\label{Lphot} 
\end{figure*} 
%==================================================================================
%% 

\section{Ionization Models: component-by-component}      
\label{app_models}

\subsection{PI models for the high-ions}        
\label{app_phothigh}

Here, we model the three high-ionization components \linebreak(H--3, H--4, and H--5) in which both \OVI\ and \NV\ are positively detected and their column densities are well measured. As demonstrated in the previous section, the $N(\HI)$ limits in H--3 and H--5 are also robust. For each of these components we compute ionization fractions for different species as a function of density for a given input metallicity using the corresponding $N(\HI)$ limits as the stopping criterion for {\sc cloudy}. In order to explore possible dependence of ionization fractions on the input metallicity, models were run for a series of different input metallicities (e.g. ${\rm [X/H]} =$~$-$1.0, 0.0, 0.3, 0.7, and 1.0). We, however, find that the ionization fractions do not show any significant dependence on the input metallicity in the optically thin regime for $\rm [X/H] < 0.3$ for the entire density range of interest. For the components with derived metallicity $\rm [X/H] > 0.3$, we use the {\sc cloudy} model for which input metallicity is close to the derived metallicity. Using the model predicted ionization fractions and the observed column densities we determine loci of \OVI\ and \NV\ in the density--metallicity ($\log n_{\rm H}-{\rm [X/H]}$) plane.

In different panels of Figure~\ref{Hphot}, these loci of \OVI\ and \NV\ are shown. The curve corresponding to each ion (\OVI\ or \NV) represents the required metallicity to reproduce the observed column density as a function of $\log n_{\rm H}$ (or $\log U$). The range in $\rm [X/H]$ and in $\log n_{\rm H}$ for which both the curves overlap, represent a PI equilibrium solution since both $N(\OVI)$ and $N(\NV)$ are simultaneously reproduced.

In Table~\ref{tab_PIhigh} we summarize the PI model solutions for the three high-ionization components. Densities as estimated for different components are consistent within $\sim0.4$ dex with a median value of $n_{\rm H} \sim 10^{-4.2}$ cm$^{-3}$ (i.e. $\log U = -1.65$). The most interesting aspect of our models is to note the conservative lower limits on gas phase metallicities. All three components show super-solar metallicities with $\rm [X/H] \gtrsim 0.3$. For the H--3 and H--5 components, in which the $N(\HI)$ limits are well constrained, the lower limits turn out to be $\rm [X/H] \gtrsim 0.6$ and $\rm [X/H] \gtrsim 0.7$, respectively. These metallicities are very high and very rare for any intervening absorption system. Such super-solar metallicities, however, is common in intrinsic QSO absorbers \citep[e.g.][]{Petitjean94a,Wu10,Muzahid13} that stem from the black hole accretion disk-wind. The total hydrogen column density is in the range $\log N_{\rm H} = 17.8 - 18.6$ and the line-of-sight thickness range from 4 -- 10 kpc. Note that the PI solutions for the high-ionization components did not produce any considerable column densities for the low-ions (e.g. \CII, \SiII, \SiIII) but produced strong \CIII, \CIV, and moderate \SiIV\ column densities. Finally, we find that the PI modeling for the H--1 and H--2 components, in which \NV\ is not detected, suggest $\log n_{\rm H} < -4.1$ and $\rm [X/H] \lesssim 0.0$.

%%%%%%%%%%%%%
\begin{table*}  
\caption{Photoionization model parameters for the high-ionization absorption components.}          
\begin{tabular}{cccccccccccccc}    
\hline \hline 
Component & $\log n_{\rm H}^{a}$  &   $\log U$ &  $\gtrsim {\rm [X/H]}^{a}$ &   $>\log N_{\rm H}$ &  $>L_{\rm los}$ &   $\log T$    &  \multicolumn{7}{c}{Predicted $\log N$ (cm$^{-2}$)} \\ \cline{8-14}   
      &  (cm$^{-3}$)          &            &                 &    (cm$^{-2}$)     &  (kpc)              &    (K)        &  \NII  &  \SiII\ & \SiIII\ & \CII\ & \CIII & \CIV\ &  \SIV\ \\ 
\hline 
(1)   &       (2)             &    (3)     &     (4)         &    (5)             &  (6)                &    (7)        &  (8)    &  (9)    & (10)  &  (11) & (12)  &  (13) & (14) \\         
\hline   
H--3   &   $[-4.25$, $-4.15]$, ($-4.20$)  & $[-1.70$, $-1.60]$ &    0.60, (0.8)  &   18.1   &     6   & 3.7  &  11.8  &   10.8  &  12.2   & 12.9  & 14.7  &  14.8 &  13.2  \\   
H--4   &   $[-4.05$, $-3.85]$, ($-3.95$)  & $[-2.00$, $-1.80]$ &    0.30, (0.6)  &   18.6   &    10   & 4.0  &  12.3  &   11.4  &  12.8   & 13.3  & 15.0  &  15.1 &  13.7  \\    
H--5   &   $[-4.35$, $-4.25]$, ($-4.30$)  & $[-1.60$, $-1.50]$ &    0.70, (0.9)  &   17.8   &     4   & 4.5  &  11.6  &   10.9  &  12.1   & 12.9  & 14.6  &  14.6 &  13.1  \\     
\hline  
\label{tab_PIhigh}    
\end{tabular}       
~\\ 
Note-- $^{a}$Values in the parenthesis are used to compute column densities in columns 8 through 14.      
\end{table*}  
%%%%%%%%%%%%% 

%%%%%%%%%%%%%
\begin{table*} 
\begin{center}  
\caption{Photoionization model parameters for the low-ionization absorption components.}          
\begin{tabular}{cccccccccccccc}    
\hline \hline 
Component & $\log n_{\rm H}$  &   $\log U$ &  $\gtrsim {\rm [X/H]}^{a}$ &   $>\log N_{\rm H}$ &  $>L_{\rm los}$ &   $\log T$    &  \multicolumn{7}{c}{Predicted $\log N$ (cm$^{-2}$)} \\ \cline{8-14}   
      &  (cm$^{-3}$)          &            &                 &    (cm$^{-2}$)     &  (kpc)              &    (K)        &  \NII  &  \SiII\ & \SiIII\ & \CII\ & \CIII & \CaII\ &  \FeII\ \\ 
\hline 
(1)   &       (2)             &    (3)     &     (4)         &    (5)             &  (6)                &    (7)        &  (8)    &  (9)    & (10)  &  (11) & (12)  &  (13) & (14) \\         
\hline   
L--1   &   $-2.40$  & $-3.45$ &   $-2.00$  &   19.7   &     4   & 4.1  &  13.3  &   12.9  &  12.8   & 13.9$^{a}$  & 13.8  &  11.1 &  12.6  \\   
L--2   &   $-2.60$  & $-3.25$ &   $-1.30$  &   19.8   &     8   & 4.1  &  14.1  &   13.6  &  13.8   & 14.5      & 14.8  &  11.7 &  13.1  \\    
L--3   &   $-2.50$  & $-3.35$ &   $-1.40$  &   19.7   &     5   & 4.1  &  13.9  &   13.4  &  13.5   & 14.3      & 14.5  &  11.6 &  13.0  \\     
\hline   
\label{tab_PIlow}    
\end{tabular} 
~\\ 
Note-- $^{a}$\CII\ is used for constraining the model.         
\end{center}    
\end{table*}  
%%%%%%%%%%%%% 

\subsection{PI models for the low-ions}        
\label{app_photlow}   

In Section~\ref{sec_absana} we have demonstrated that the kinematics of \OVI\ and \MgII\ absorption are significantly different which indicate the multiphase structure of the absorber. In the previous section, we have shown that the gas phase that gives rise to \OVI\ and \NV\ under PI equilibrium cannot produce any considerable low-ions column densities. This further reinforces the idea of multiphase nature of the absorbing gas. To constrain the ionization parameters and chemical abundances in the low-ionization absorption components we have used the observed column densities of $N(\MgI)$ and $N(\MgII)$ and the upper limits on $N(\HI)$. Note that, as \MgI\ is not detected, $N(\CII)$ is used for constraining the model parameters for the L--1 component.   
%%% 

PI models for the low-ionization components L--1, L--2, and L--3 are shown in Figure~\ref{Lphot} and the model parameters are summarized in Table~\ref{tab_PIlow}. The densities in all these components are very similar (i.e. $n_{\rm H}\sim 10^{-2.5\pm0.1}$ ~cm$^{-3}$). The metallicity in the L--1 component is ${\rm [X/H]} \gtrsim -2$, but it is roughly $\sim0.7$ dex higher for the L--2 and L--3 components. It is interesting to note that while for the high-ionization components with detected \NV, data do not allow a metallicity of $\rm [X/H] < 0.3$, the low-ionization components can have significantly lower metallicities. However the metallicity need not necessarily be this low. Here, we have placed a very conservative lower limit on metallicity. The total hydrogen column densities in these components (i.e. $\log N_{\rm H} \sim 19.7-19.8$) are $\gtrsim10$ times higher than what we have found for the high-ionization components (see Table~\ref{tab_PIhigh}). The line-of-sight thicknesses of 4--8 kpc are, however, comparable to those of the high-ionization components. Note that our low-ionization phase solutions do not produce any appreciable amount of \SiIV, \CIV\ or any other higher ions (i.e. $\log N < 13.0$). The model predicted column densities for different low-ions are presented in columns 8 through 14 of Table~\ref{tab_PIlow}. We use these column densities to generate synthetic profiles using {\sc vpfit} assuming pure non-thermal broadening. The synthetic profiles are shown in magenta curves in Figure~\ref{vplot}. Note that these synthetic profiles explain the observed data adequately for \CII, \NII, \SiII, and \SiIII. Our models, however, predict 0.3 and 0.5 dex higher \CaII\ column densities in L--2 and L--4 components, respectively. It is known that Ca easily depletes on to dust. To be consistent with the data our PI model solutions would require a maximum of 0.5 dex depletion for calcium. Further, we note that our model predicted $N(\FeII)$ values are 0.4 dex higher as compared to the measured values in the L--2 and L--3 components. This can be understood in terms of $\alpha$-enhancement. Note that, although we have built very reasonable PI models, the metallicity of the low-ionization components (L--2 and L--3 in particular) are not very well constrained due to the lack of higher order Lyman series lines. Spectral coverage of higher order Lyman-series lines down to the Lyman limit will be very useful in order to estimate metallicities in different low-ionization components accurately. Nevertheless, the metallicity of the L--1 component cannot be considerably higher than $\rm [X/H] = -2.0$, in order to explain the red wing of the \lya\ profile.

%% 
%==================================================================================
\begin{figure*} 
\centerline{\vbox{
\centerline{\hbox{ 
\includegraphics[width=0.33\textwidth,angle=00]{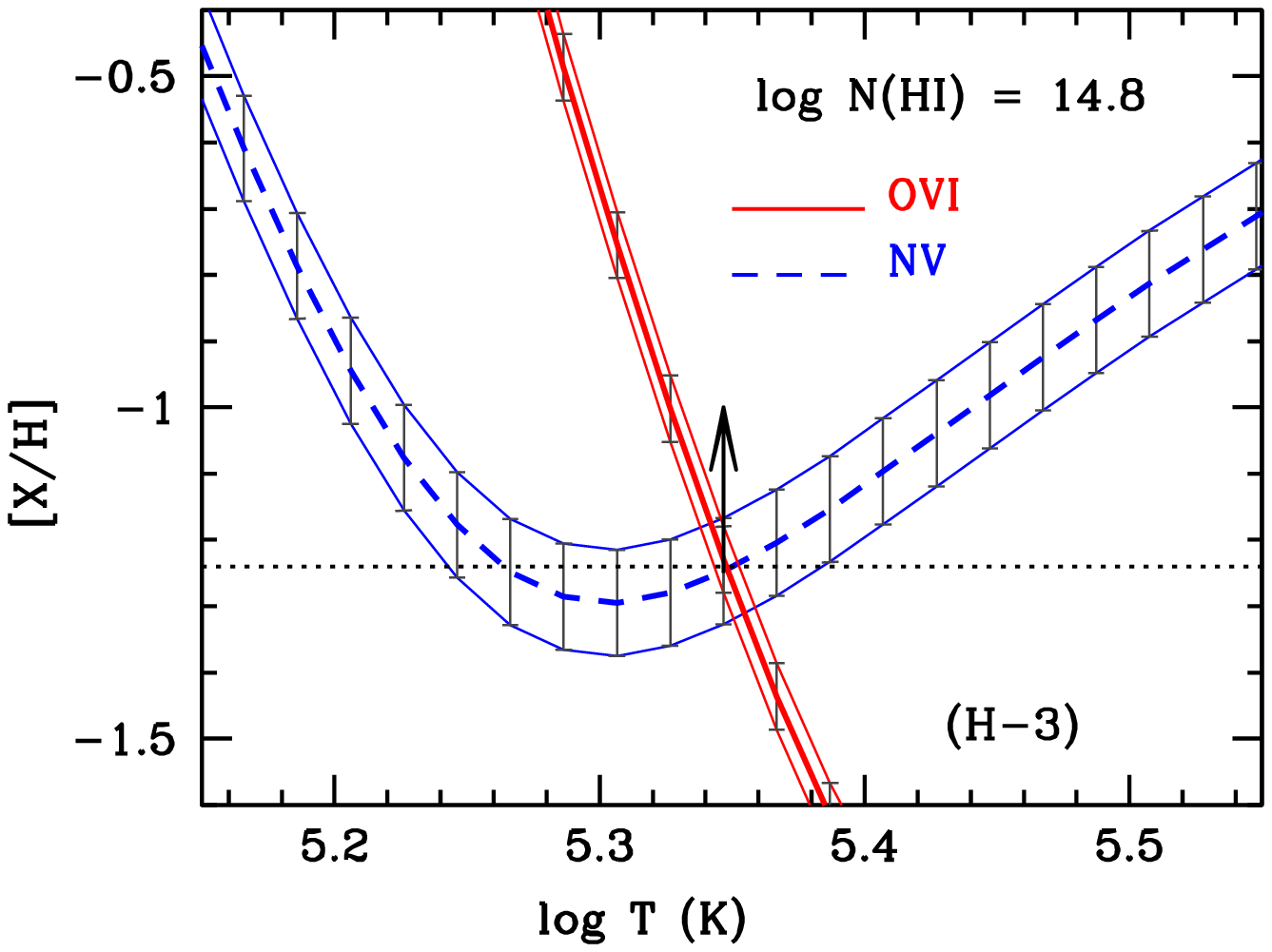}
\includegraphics[width=0.33\textwidth,angle=00]{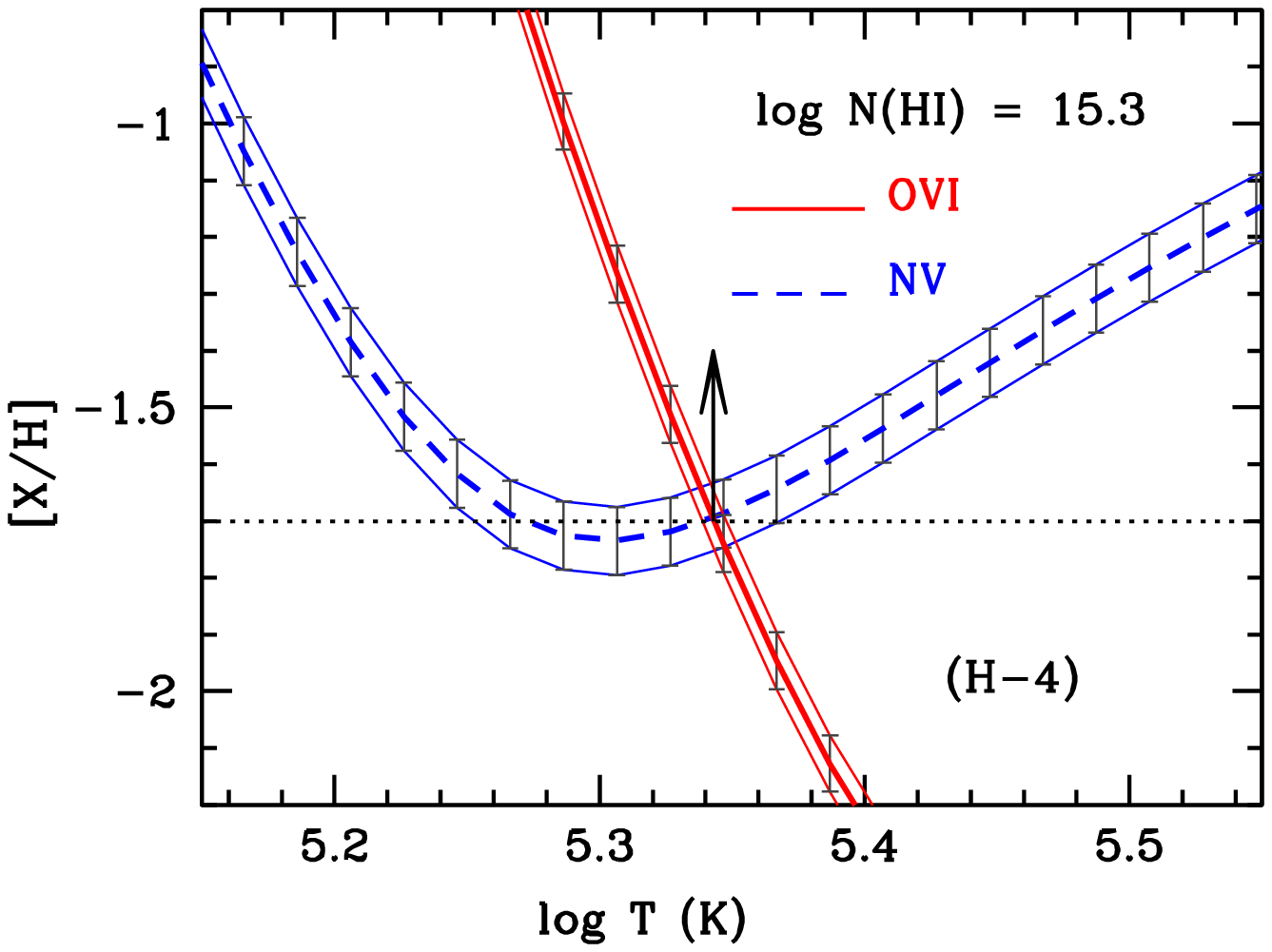}  
\includegraphics[width=0.33\textwidth,angle=00]{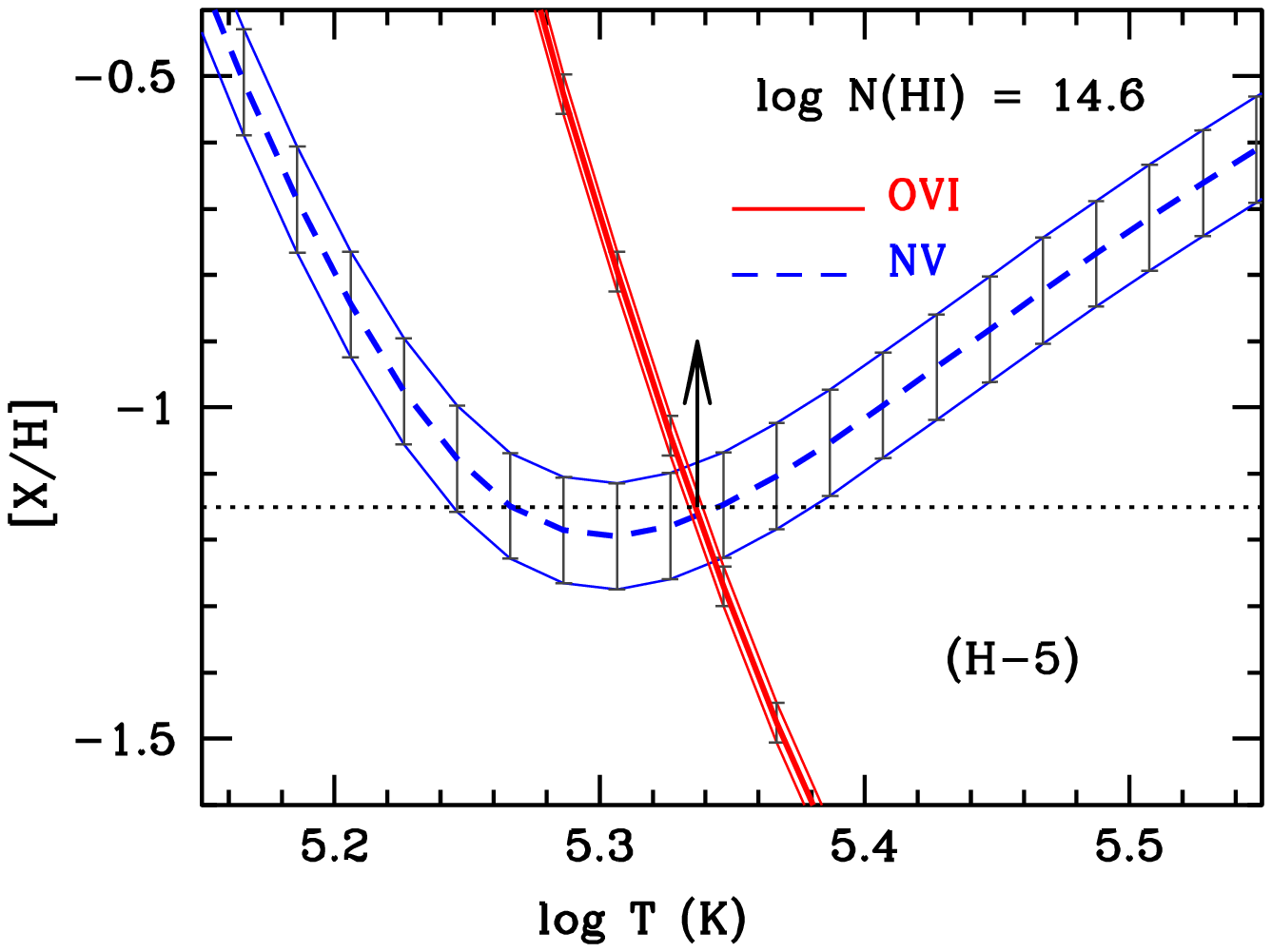} 
}}
}}   
\vskip-0.3cm  
\caption{CIE models for the high-ionization absorption components, H--3 (left panel),  
H--4 (middle panel), and H--5 (right panel). In each panel, different curves represent 
required metallicities to reproduce observed column densities for different high-ions 
(\OVI\ and \NV) as a function of gas temperature. The $N(\HI)$ values assumed by the models 
are mentioned in the plot (see Table~\ref{tab_highion}). The region in which the two 
curves overlap represents a CIE solution. It is evident that for all three components the 
required gas temperature is $\log T > 5.3$, in order to reproduce the observed $N(\OVI)$ and 
$N(\NV)$ simultaneously. The arrow indicates the lower limit on metallicity in each panel.               
}            
\label{HCIE} 
\end{figure*} 
%==================================================================================
%%  

%% 
%==================================================================================
\begin{figure*} 
\centerline{\vbox{
\centerline{\hbox{ 
\includegraphics[width=0.33\textwidth,angle=00]{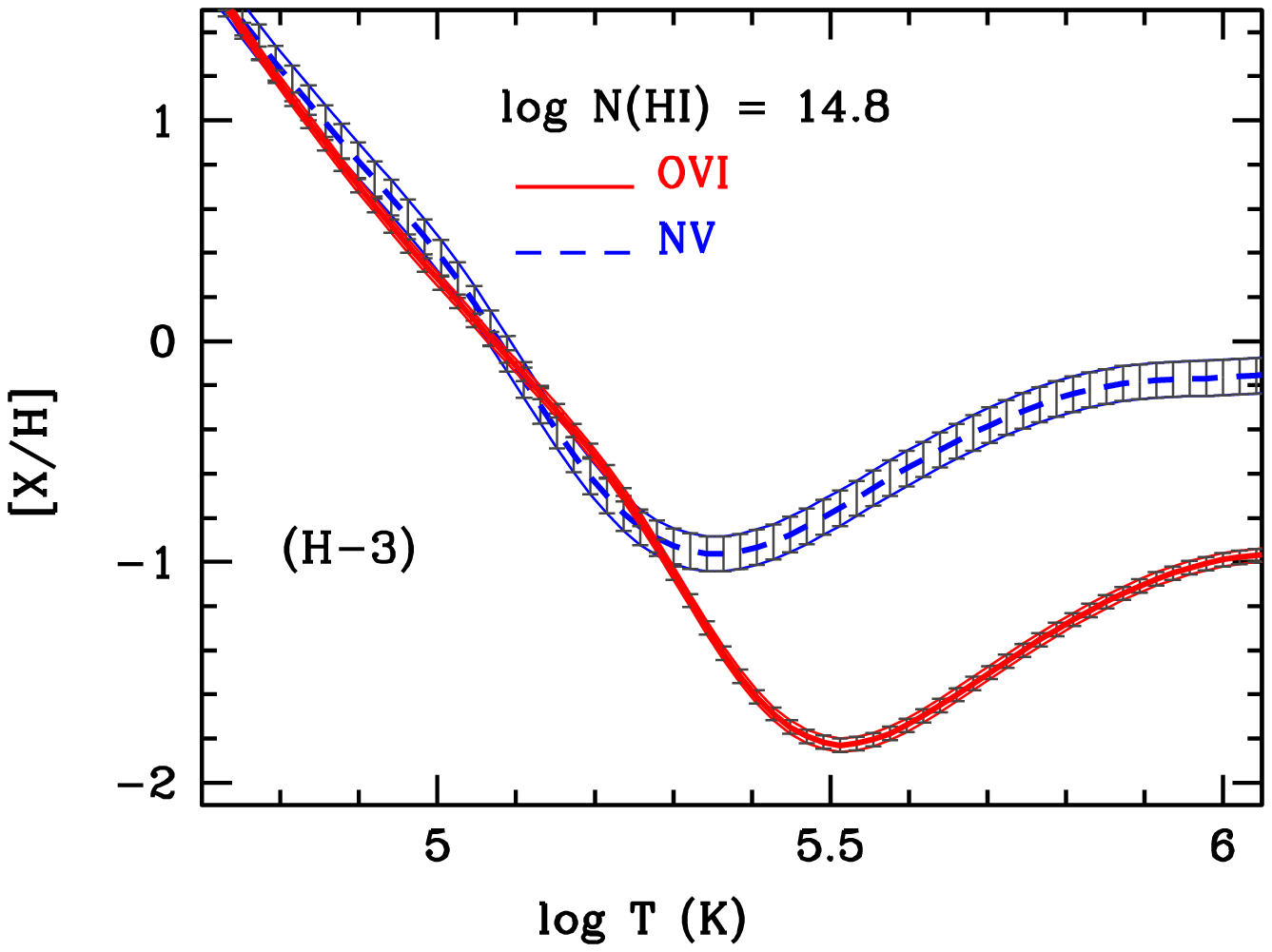}
\includegraphics[width=0.33\textwidth,angle=00]{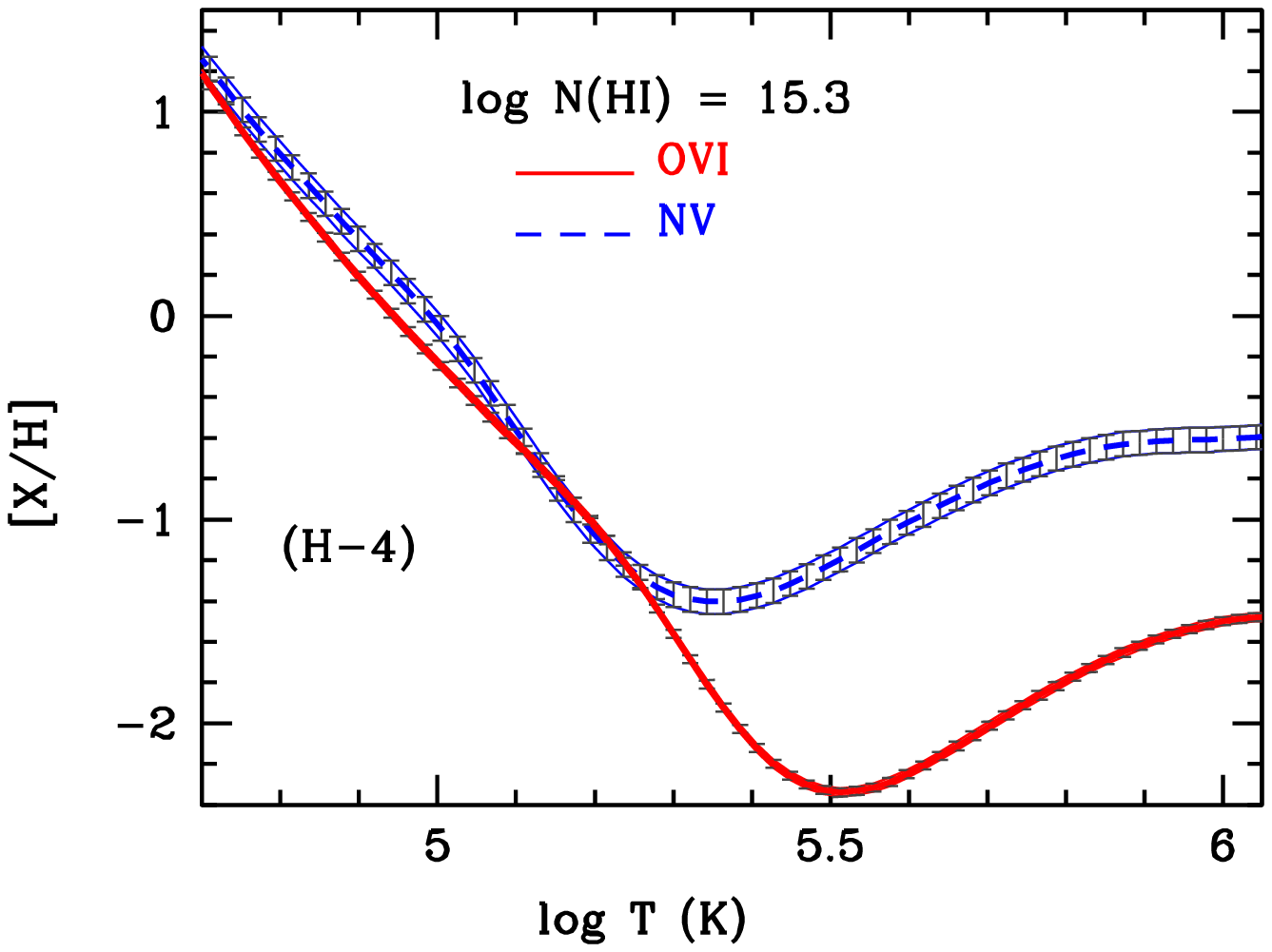}  
\includegraphics[width=0.33\textwidth,angle=00]{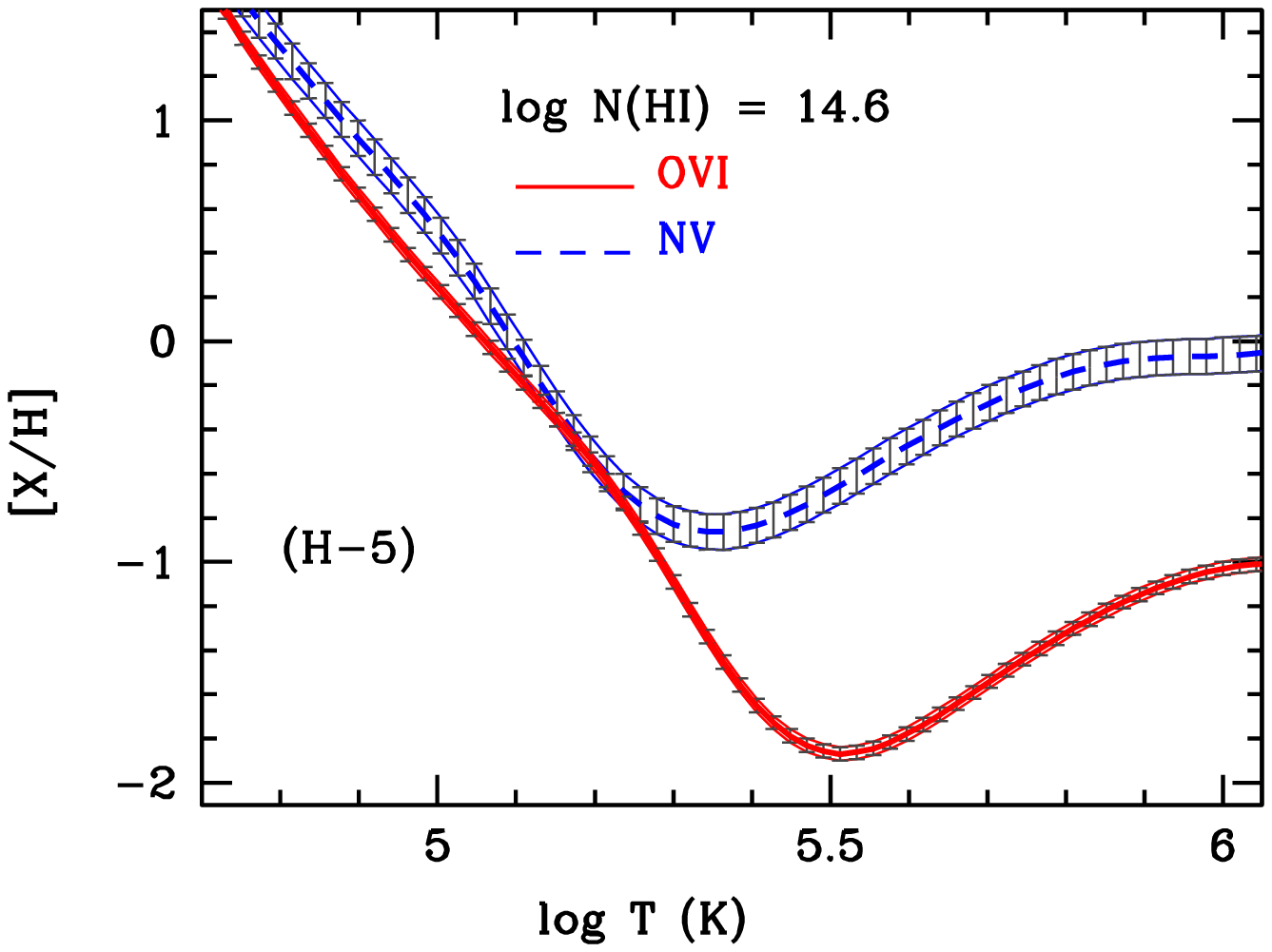} 
}}
}}   
\vskip-0.3cm  
\caption{Non-CIE isobaric cooling models for the high-ionization absorption components. Everything 
else is same as in Figure~\ref{HCIE}. It is evident that the observed $N(\OVI)$ and $N(\NV)$ can be 
reproduced simultaneously for gas temperature in the range $\log T \simeq 5.1 - 5.2$.      
}            
\label{HnCIE_isobaric}  
\end{figure*} 
%==================================================================================
%%  

%% 
%==================================================================================
\begin{figure*} 
\centerline{\vbox{
\centerline{\hbox{ 
\includegraphics[width=0.33\textwidth,angle=00]{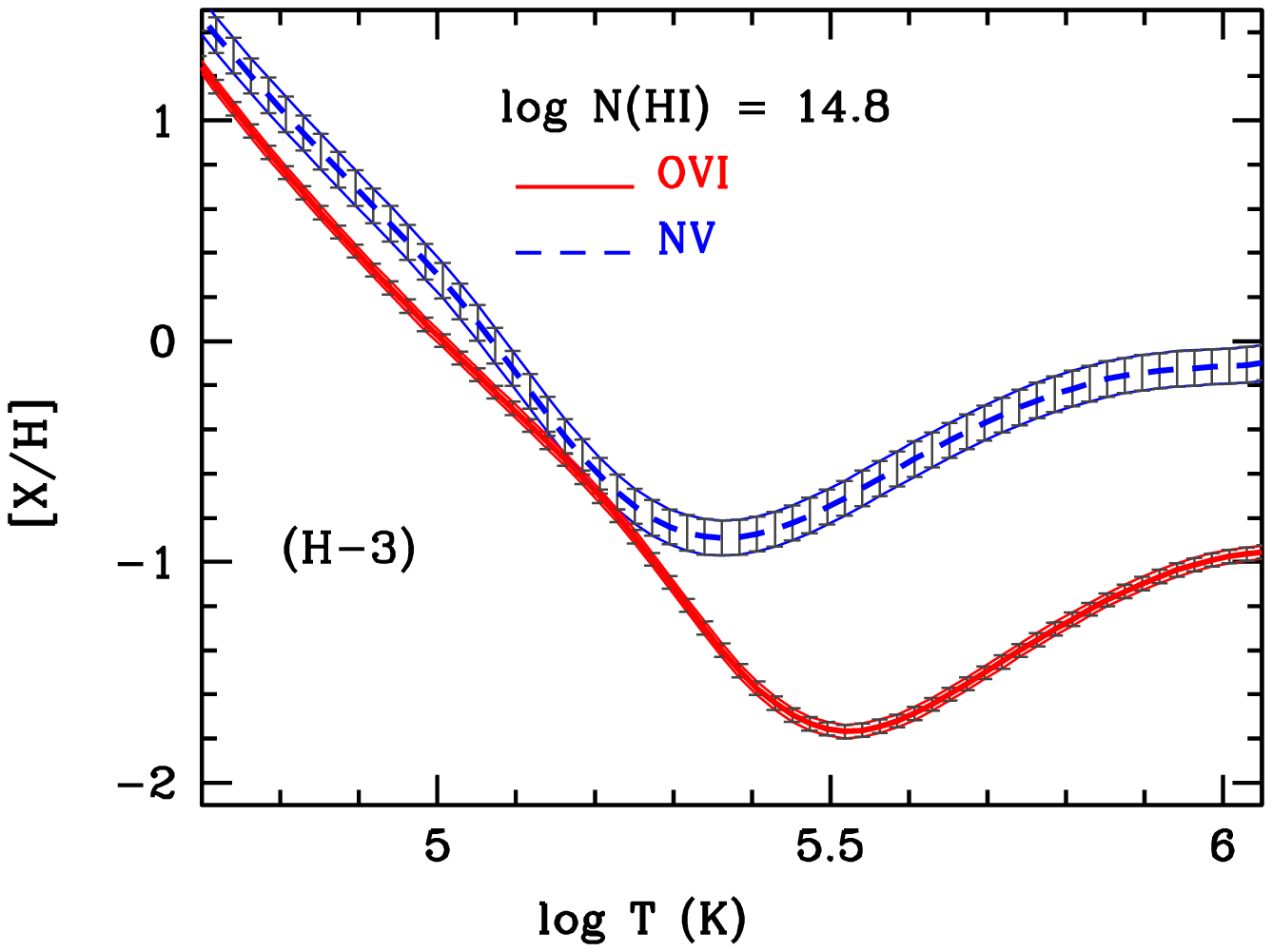}
\includegraphics[width=0.33\textwidth,angle=00]{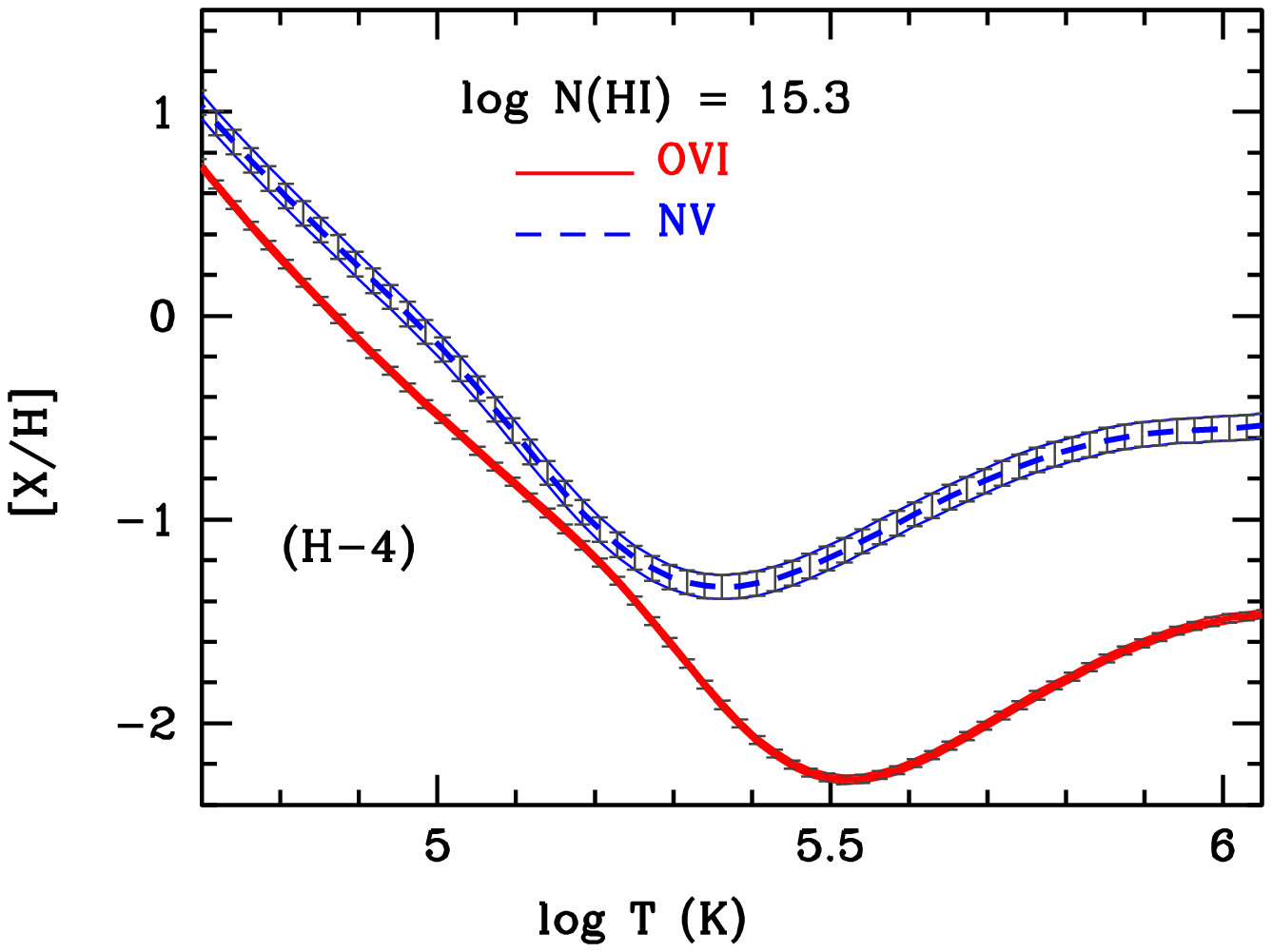}  
\includegraphics[width=0.33\textwidth,angle=00]{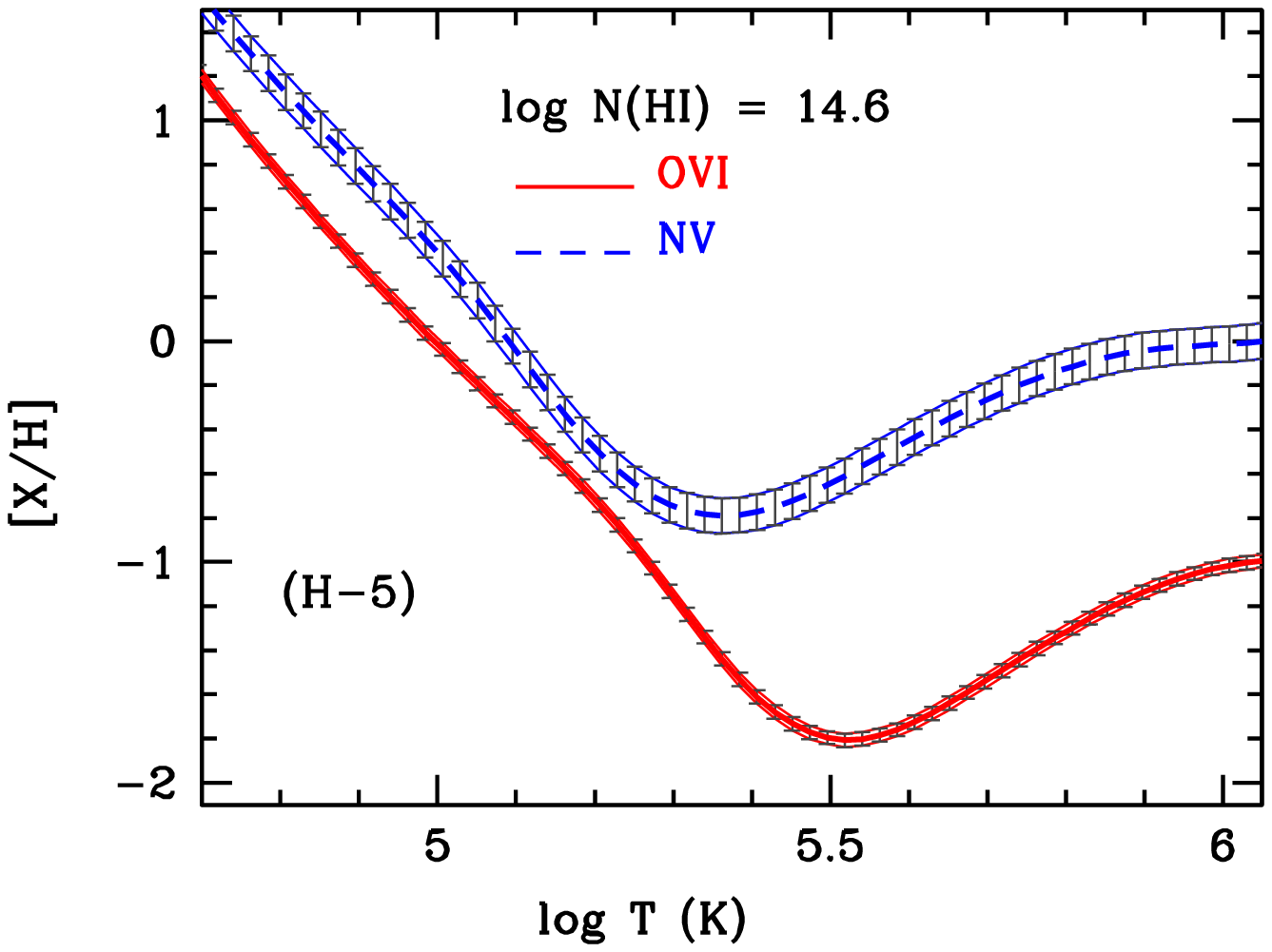} 
}}
}}  
\vskip-0.3cm  
\caption{Non-CIE isochoric cooling models for the high-ionization absorption components. Everything 
else is same as in Figure~\ref{HCIE}. It is apparent that the observed $N(\OVI)$ and $N(\NV)$ are not 
reproduced simultaneously for any of the components for the entire parameter space indicating no 
non-CIE solution.  
}            
\label{HnCIE_isochoric} 
\end{figure*} 
%==================================================================================
%%  

\subsection{CIE models for the high-ions}          
\label{app_CIE}   

In this section we will explore the viability of CIE models for the  high-ions. CIE models for the H--3, H--4, and H--5 components, in which \NV\ is positively detected, are shown in Figure~\ref{HCIE}. We adopt the CIE models of \citet{Gnat07}. In their models, equilibrium and non-equilibrium cooling efficiencies and ionization states for low-density radiatively cooling gas are computed under the assumption that the gas is dust-free, optically thin, and subject to no external radiation field. In each panels of Figure~\ref{HCIE}, we plot the metallicity required to produce the observed $N(\OVI)$ and $N(\NV)$ as a function of gas temperature ($\log T$) under CIE conditions. It is apparent from the figure that the observed $N(\OVI)$ and $N(\NV)$ are simultaneously explained for gas temperature in the range, $\log T \gtrsim 5.3 - 5.4$. The lower limit on metallicity for the H--3 and H--5 components, in which $N(\HI)$ values are robustly constrained, turns out to be $\rm [X/H] > -1.2$ and for the H--4 component, $\rm [X/H] > -1.7$. Therefore, in contrast to the PI models, the gas-phase metallicity can be significantly lower under CIE conditions provided the gas temperature is $\log T \gtrsim 5.3$. Such a high temperature corresponds to a \HI\ Doppler parameter of $b(\HI) \gtrsim 57$ \kms. However, this large $b(\HI)$ value, as suggested by the CIE models, is not supported by the COS data. As mentioned earlier in Appendix~\ref{app_NHIestimate} a free fit to the non-black absorption seen in the \lyb\ profile at $v \sim +100$ \kms\ (i.e. the H--3 component) gives $b(\HI) = 28\pm4$ \kms. This is half of the $b(\HI)$ value required by the CIE models. Furthermore, we find that the CIE solutions do not produce any considerable $N(\CIV)$ in any of the components. The maximum $N(\CIV)$ predicted by our CIE model is $\log N = 13.9$. However, in the low resolution $HST/$FOS spectrum, \citet{Chen01} have detected a very strong \CIV\ absorption from this system with $W_{r}= 1.70$ \AA\footnote{Note that we attempted to observed the \CIV\ absorption using the G185M grating. However, the resultant spectrum fails to put any useful constraints to our models due to poor $S/N$ ($<2$ per resolution element).}. For comparison, the measured rest-frame equivalent width of \OVI~$\lambda 1031$ absorption is $W_{r}= 1.02\pm0.02$ \AA. Therefore, the \CIV\ is even stronger than the \OVI\ in this system, since the $f\lambda$ value for the \CIV~$\lambda1548$ transition is $\gtrsim2$ times higher than that of \OVI~$\lambda 1031$. Nonetheless, CIE models fail to reproduce strong enough \CIV\ along with \NV\ and \OVI. These facts indicate that a pure CIE model is {\sl not} viable. Photoionization models (Appendix~\ref{app_phothigh}), however, can reproduce strong \CIV\ along with \NV\ and \OVI\ from a single gas phase.

\subsection{Non-CIE models for the high ions}          
\label{app_nCIE}

In this section, we investigate if non-CIE models can explain the high-ionization absorption components. Note that non-equilibrium effects become important in metal-rich environments when cooling efficiencies are increased and gas can cool faster than it recombines. Here, we consider two different non-CIE models of \citet{Gnat07}, namely, isochoric (constant density) and isobaric (constant pressure) models. As discussed by the authors, isobaric cooling is less rapid compared to isochoric cooling and departures from equilibrium are somewhat smaller. Moreover, cooling that is initially isobaric eventually transforms to isochoric cooling (see their Figure~1).

In Figure~\ref{HnCIE_isobaric} we show non-CIE isobaric cooling models for the H--3, H--4, and H--5 components. Clearly, these models suggest a gas temperature (e.g. $\log T \simeq 5.1 - 5.2$) which is only slightly lower than that predicted by the pure CIE models (see Figure~\ref{HCIE}). Note that for the H--3 component, the observed $N(\OVI)/N(\NV)$ ratio is consistent for a large range in gas temperature (i.e. $\log T \sim 4.7 - 5.3$). The lower limit on metallicity for $\log T \sim5.3$ is $\rm [X/H] > -1.0$. The required metallicity sharply increases as the gas temperature decreases. For example, any temperature lower than $10^{5.1}$~K would require a super-solar metallicity. We note that for $\log T < 5.1$ our model overproduces $N(\SIV)$ by $\sim0.4$ dex. Moreover, a temperature of $\log T \lesssim {4.8}$ overproduces $N(\SiIII)$ by $\sim2$ dex. We, thus, rule out the possibility of a gas temperature  lower than $10^{5.1}$~K. Unlike CIE models, an isobaric cooling model solution with $\rm [X/H] \sim 0.0$ and $\log T \sim 5.1$ can produce a large \CIV\ column density (e.g. $\log N> 15.1$). Nonetheless, as discussed in the previous section, the $b$-parameter required for such a high temperature (i.e. $b(\HI) > 46$~\kms) is not supported by the \lyb\ profile.

In Figure~\ref{HnCIE_isochoric} we show non-CIE isochoric cooling models for the H--3, H--4, and H--5 components. Note that we do not find any possible solution that could explain the observed $N(\OVI)$ and $N(\NV)$ simultaneously within a $1\sigma$  measurement uncertainty. This clearly indicates that a non-CIE isochoric cooling model is not feasible. However, within $2\sigma$ allowed uncertainties we can have solutions which can be, again, ruled out by the arguments that were used previously to rule out the isobaric models.

\section{Equations for outflow energetics}  
\label{app_equations}
In all the following equations: $r$ is the thin-shell radius, $\Delta r$ is the cloud thickness, 
$v_w$ is the outflow speed, $T$ is the cloud temperature, $N_{\rm H}$ is the total hydrogen 
column density, $C_{\Omega}$ is the global covering factor, $C_{f}$ is the local covering 
factor, $t_{\rm flow}$ is the flow time, $t_{\rm sound}$ is the sound crossing time, $M_{\rm out}$ 
is the mass carried by the thin-shell, $\dot{M}_{\rm out}$ is the mass-flow rate, and $\dot{E}_{k}$ 
is the kinetic luminosity. All other symbols have their usual meanings. We refer the reader \cite{Rupke05} 
for detailed derivations of the Equations~\ref{mout}, \ref{mdotout}, and \ref{edotout}.

\begin{equation}   
\footnotesize   
t_{\rm flow} = 5 \times10^{8}~ {\rm yrs} \left(\frac{r}{100 ~{\rm kpc}}\right)  
\left(\frac{v_w}{200 ~{\rm km s^{-1}}}\right)^{-1} 
\end{equation}   

\begin{equation}    
\footnotesize   
t_{\rm sound} = 8 \times10^{8}~ {\rm yrs} \left(\frac{\Delta r}{10~{\rm kpc}}\right)  
\left(\frac{T}{10^{4} ~{\rm K}}\right)^{-\frac{1}{2}}    
\end{equation}

\begin{equation}   
\footnotesize   
\begin{split} 
M_{\rm out} &  = 4 \pi \mu m_{p} C_{\Omega} C_f N_{\rm H} r^2    
             =  5.6\times10^{10} M_{\odot} \\  
& \times \left(\frac{C_{\Omega}}{0.4} C_f\right)  
\left(\frac{N_{\rm H}}{10^{20} ~ {\rm cm^{-2}}} \right)    
\left(\frac{r}{100 ~ {\rm kpc}}\right)^2      
\end{split} 
\label{mout}  
\end{equation}

\begin{equation}
\footnotesize   
\begin{split} 
{\dot M}_{\rm out} &  =  8 \pi \mu m_{p} C_\Omega C_f N_{\rm H} r v_w       
                     = 230 ~ M_{\odot} ~{\rm yr^{-1}}  \\ 
& \times \left(\frac{C_\Omega}{0.4} C_f\right)      
\left(\frac{N_{\rm H}}{10^{20} ~ {\rm cm^{-2}}} \right)    
\left(\frac{r}{100 ~ {\rm kpc}}\right)             
\left(\frac{v_w}{200 ~ {\rm km s^{-1}}}\right)  
\end{split} 
\label{mdotout}   
\end{equation}

\begin{equation}
\footnotesize   
\begin{split} 
{\dot E_{k}} &  = \frac{1}{2}{\dot M} {v_w}^2  = 4 \pi \mu m_{p} C_\Omega C_f N_{\rm H} r {v_w}^3     
               = 2.9\times10^{42} ~{\rm erg~s^{-1}}   \\ 
& \times \left(\frac{C_\Omega}{0.4} C_f\right)       
\left(\frac{N_{\rm H}}{10^{20} ~ {\rm cm^{-2}}} \right)     
\left(\frac{r}{100 ~ {\rm kpc}}\right)        
\left(\frac{v_w}{200 ~ {\rm km s^{-1}}}\right)^3       
\end{split} 
\label{edotout}   
\end{equation}

%----------------- Bibliography and bibfile  ----------------------  
\def\aj{AJ}%
\def\actaa{Acta Astron.}%
\def\araa{ARA\&A}%
\def\apj{ApJ}%
\def\apjl{ApJ}%
\def\apjs{ApJS}%
\def\ao{Appl.~Opt.}%
\def\apss{Ap\&SS}%
\def\aap{A\&A}%
\def\aapr{A\&A~Rev.}%
\def\aaps{A\&AS}%
\def\azh{AZh}%
\def\baas{BAAS}%
\def\bac{Bull. astr. Inst. Czechosl.}%
\def\caa{Chinese Astron. Astrophys.}%
\def\cjaa{Chinese J. Astron. Astrophys.}%
\def\icarus{Icarus}%
\def\jcap{J. Cosmology Astropart. Phys.}%
\def\jrasc{JRASC}%
\def\mnras{MNRAS}%
\def\memras{MmRAS}%
\def\na{New A}%
\def\nar{New A Rev.}%
\def\pasa{PASA}%
\def\pra{Phys.~Rev.~A}%
\def\prb{Phys.~Rev.~B}%
\def\prc{Phys.~Rev.~C}%
\def\prd{Phys.~Rev.~D}%
\def\pre{Phys.~Rev.~E}%
\def\prl{Phys.~Rev.~Lett.}%
\def\pasp{PASP}%
\def\pasj{PASJ}%
\def\qjras{QJRAS}%
\def\rmxaa{Rev. Mexicana Astron. Astrofis.}%
\def\skytel{S\&T}%
\def\solphys{Sol.~Phys.}%
\def\sovast{Soviet~Ast.}%
\def\ssr{Space~Sci.~Rev.}%
\def\zap{ZAp}%
\def\nat{Nature}%
\def\iaucirc{IAU~Circ.}%
\def\aplett{Astrophys.~Lett.}%
\def\apspr{Astrophys.~Space~Phys.~Res.}%
\def\bain{Bull.~Astron.~Inst.~Netherlands}%
\def\fcp{Fund.~Cosmic~Phys.}%
\def\gca{Geochim.~Cosmochim.~Acta}%
\def\grl{Geophys.~Res.~Lett.}%
\def\jcp{J.~Chem.~Phys.}%
\def\jgr{J.~Geophys.~Res.}%
\def\jqsrt{J.~Quant.~Spec.~Radiat.~Transf.}%
\def\memsai{Mem.~Soc.~Astron.~Italiana}%
\def\nphysa{Nucl.~Phys.~A}%
\def\physrep{Phys.~Rep.}%
\def\physscr{Phys.~Scr}%
\def\planss{Planet.~Space~Sci.}%
\def\procspie{Proc.~SPIE}%
\let\astap=\aap
\let\apjlett=\apjl
\let\apjsupp=\apjs
\let\applopt=\ao
\bibliographystyle{apj}
\bibliography{mybib}
%-------------------------------------------------------------------   
%-------------------------------------------------------------------   

\end{document}